\documentclass[twocolumn,hyperpdf,amsmath,amssymb,aps,prd,10pt,superscriptaddress,nofootinbib,noeprint,preprintnumbers,floatfix]{revtex4-1}

\usepackage{graphicx, color}

\usepackage[letterspace=-10]{microtype} 

\usepackage{bm, amsmath, amsfonts, amssymb,xfrac}
\usepackage{multirow, tabularx, dcolumn}
\usepackage{mathtools,leftidx, braket, slashed, cancel, bigdelim}
\usepackage{blkarray}
\usepackage[figures]{rotating}

\usepackage[utf8]{inputenc} 
\usepackage{hyperref}
\definecolor{jlab_red}{RGB}{192,39,45}
\definecolor{jlab_orange}{RGB}{249,102,0}
\definecolor{jlab_blue}{RGB}{47,122,121}
\definecolor{jlab_green}{RGB}{65,125,10}
\definecolor{jlab_gray}{gray}{0.6}

\newcommand{\etaOctet}{     \eta^{\scriptscriptstyle\mathbf{8}}}
\newcommand{\etaSinglet}{   \eta^{\scriptscriptstyle\mathbf{1}}}
\newcommand{\omegaOctet}{   \omega^{\scriptscriptstyle\mathbf{8}}}
\newcommand{\omegaSinglet}{ \omega^{\scriptscriptstyle\mathbf{1}}}
\newcommand{\fOneOctet}{    f_1^{\scriptscriptstyle\mathbf{8}}}
\newcommand{\fOneSinglet}{  f_1^{\scriptscriptstyle\mathbf{1}}}

\newcommand{\fZeroSinglet}{  f_0^{\scriptscriptstyle\mathbf{1}}}

\newcommand{\hOneOctet}{    h_1^{\scriptscriptstyle\mathbf{8}}}
\newcommand{\hOneSinglet}{  h_1^{\scriptscriptstyle\mathbf{1}}}

\newcommand{\threeSone}{\prescript{3\!}{}{S}_1} 

\newcommand{\onePone}{\prescript{1\!}{}{P}_1} 

\newcommand{\threePone}{\prescript{3\!}{}{P}_1}
\newcommand{\threePtwo}{\prescript{3\!}{}{P}_2}  
\newcommand{\fivePone}{\prescript{5\!}{}{P}_1} 
\newcommand{\fivePtwo}{\prescript{5\!}{}{P}_2} 
\newcommand{\fivePthree}{\prescript{5\!}{}{P}_3} 
\newcommand{\threeDone}{\prescript{3\!}{}{D}_1}
\newcommand{\threeDtwo}{\prescript{3\!}{}{D}_2}
\newcommand{\threeDthree}{\prescript{3\!}{}{D}_3}

\newcommand{\oneDtwo}{\prescript{1\!}{}{D}_2}

\newcommand{\threeFtwo}{\prescript{3\!}{}{F}_2}
\newcommand{\threeFthree}{\prescript{3\!}{}{F}_3}

\newcommand{\oneFthree}{\prescript{1\!}{}{F}_3} 
  
\newcommand{\fiveFthree}{\prescript{5\!}{}{F}_3}  
\hypersetup{%
pdftitle = {TITLE},
pdfsubject = {},
pdfkeywords = {},
pdfauthor = {Hadron Spectrum Collaboration},
colorlinks = {true},
filecolor = {black},
linkcolor = {jlab_blue},
menucolor = {black},
citecolor = {jlab_green},
urlcolor = {jlab_green},
}{}

\begin{document}

\preprint{JLAB-THY-24-4079}
\title{Coupled-channel $J^{--}$ meson resonances from lattice QCD}
\author{Jozef~J.~Dudek}
\email{dudek@jlab.org}
\affiliation{Department of Physics, College of William and Mary, Williamsburg, VA 23187, USA}
\affiliation{\lsstyle Thomas Jefferson National Accelerator Facility, 12000 Jefferson Avenue, Newport News, VA 23606, USA}
\author{Christopher~T.~Johnson}
\affiliation{Department of Physics, College of William and Mary, Williamsburg, VA 23187, USA}
\collaboration{for the Hadron Spectrum Collaboration}
\date{ \today }
\begin{abstract}
We extend an earlier calculation within lattice QCD of excited light meson resonances with $J^{PC}=1^{--}, 2^{--}, 3^{--}$ at the SU(3) flavor point in the \emph{singlet} representation, by considering the \emph{octet} representation. In this case the resonances appear in coupled-channel amplitudes, which we determine, establishing the relative strength of pseudoscalar-pseudoscalar to pseudoscalar-vector decays. Combining the new octet results with the prior results for the singlet, we perform a plausible extrapolation to the physical quark mass, and compare to experimental $\rho^\star_J, K^\star_J, \omega^\star_J$ and $\phi^\star_J$ resonances. 
\end{abstract}
\maketitle

\section{Introduction}
  \label{sec:Intro}
 
Beyond the well-established lightest vector mesons, the $\rho, \omega, \phi$ and $K^*$, the spectrum of $J^{PC} = J^{--}$ hadron resonances is quite poorly understood, with even the number of such states below 2 GeV being unclear experimentally. 
The ``OZI-rule'', an empirical observation that in many cases processes described by diagrams in which quark lines are completely disconnected are heavily suppressed relative to those where all lines are connected, is often taken as exact, and based upon it, assignments of flavor structure for resonances made. However, this ``rule'' is not a demonstrated feature of QCD in all channels, and its use may be introducing a systematic error into our understanding of the hadron spectrum. Establishing the validity of rules like this one is a major aim of modern hadron spectroscopy~\cite{Shepherd:2016dni}. 
An ongoing mystery is why there are no experimental candidate states for light $2^{--}$ resonances outside the strange sector. These states cannot decay into pairs of pseudoscalars, but can decay into pseudoscalar-vector final states which are quite well explored experimentally.

An initial exploration of these excited $J^{--}$ meson resonances in first-principles QCD appeared in Ref.~\cite{Johnson:2020ilc}. Lattice gauge-field configurations having three flavors of dynamical quarks all set equal to the physical strange quark mass were used to compute correlation functions from which the discrete finite-volume spectrum in five different spatial volumes were extracted.
These energy levels, corresponding to the \emph{singlet} ($\mathbf{1}$) representation of exact SU(3) flavor symmetry were used to constrain elastic pseudoscalar-vector scattering amplitudes through the L\"uscher finite-volume  formalism~\cite{Luscher:1985dn,Luscher:1986pf} (reviewed in Ref.~\cite{Briceno:2017max}). The resulting amplitudes were found to contain pole singularities that were identified as a narrow $3^{--}$ resonance, a broad $2^{--}$ resonance, and two overlapping $1^{--}$ resonances. 
A crude but plausible extrapolation to the physical quark masses was performed, and properties of excited $\rho_J^\star$, $\omega_J^\star$, and $\phi_J^\star$ resonances predicted, \emph{assuming} an exact implementation of the ``OZI-rule''.

This paper presents an extension of the calculation reported on in Ref.~\cite{Johnson:2020ilc} to consider, in addition, the independent \emph{octet} ($\mathbf{8}$) representation of SU(3) flavor. $J^P = 1^-, 3^-$ octet states have pseudoscalar-pseudoscalar decays in addition to pseudoscalar-vector decays, meaning that \emph{coupled-channel} amplitudes are required to describe the relevant scattering. From the pole singularities of the determined amplitudes we will be able to extract the relative strength of these decay modes.

The ``OZI-rule'' assumptions assumed in the extrapolations of Ref.~\cite{Johnson:2020ilc} will be tested, to a certain extent, in this calculation by comparing octet resonance decays to certain final states with those for singlet resonances.
Implications for amplitude extraction in a finite-volume of the near degeneracy of the lightest stable octet and singlet vector mesons, $\omegaOctet$, $\omegaSinglet$, will be explored, establishing that the current level of statistical precision on the finite-volume energy spectrum only allows us to determine with some precision the total decay rate of octet resonances into $\etaOctet \omegaOctet$ plus $\etaOctet \omegaSinglet$, but not their relative strength.  

Extrapolations comparable to those in Ref.~\cite{Johnson:2020ilc} will be presented, relaxing somewhat assumptions about the ``OZI-rule'', with $\rho_J^\star$ and $K^\star_J$ resonances following directly from the new octet results, while $\omega_J^\star$ and $\phi_J^\star$ resonances will be predicted using a combination of the new octet and the previous singlet results. 

Novel features of the current calculation will be presented in this document, while relevant background information can be found in Ref.~\cite{Johnson:2020ilc}.

\section{Finite-Volume Spectrum}
  \label{sec:Spectrum}

For the current calculation we use the same anisotropic lattices as Ref.~\cite{Johnson:2020ilc} which have three degenerate dynamical quark flavors, tuned to approximately the physical strange quark mass, leading to a lightest pseudoscalar mass near 700 MeV~\cite{Edwards:2008ja,Lin:2008pr}. \emph{Distillation}~\cite{Peardon:2009gh} is used to compute matrices of two-point correlation functions across five lattice volumes presented in 
Table~\ref{tab:lattices}. Discussion of the stable mesons on these lattices and their dispersion relations can be found in Ref.~\cite{Johnson:2020ilc}, and we reproduce their masses here in Table~\ref{tab:masses}. These same lattices were also used to establish the presence of an exotic $J^{PC}=1^{-+}$ resonance in Ref.~\cite{Woss:2020ayi}.


\begin{table}
\renewcommand{\arraystretch}{1.2}
\begin{tabular}{c|ccccc}
$L/a_s$ 			& $14$	& $16$ 	& $18$ 	& $20$	& $24$ \\
\hline
$N_\mathrm{cfgs}$	& $397$	& $490$	& $358$	& $477$	& $499$ \\
$N_\mathrm{vecs}$	& $48$	& $64$	& $96$	& $128$	& $160$ \\
$N_\mathrm{tsrcs}$	& $16$	& $4$	& $4$	& $4$	& $1$ 
\end{tabular}
\caption{
Number of distillation vectors ($N_{\text{vecs}}$), gauge configurations ($N_{\text{cfgs}}$), and time-sources ($N_{\text{tsrcs}}$) used in computation of two-point correlation functions on each lattice volume.}
\label{tab:lattices}
\end{table}

\begin{table}
\centering\renewcommand{\arraystretch}{1.2}
\begin{tabular}[t]{ r l @{\hskip 3.0ex}|@{\hskip 3.0ex} r l } 
	$\etaOctet(0^{-+})$ & 0.1478(1) 	& $\etaSinglet(0^{-+})$ & 0.2017(11)   \\
	$\omegaOctet(1^{--})$ & 0.2154(2) 	& $\omegaSinglet(1^{--})$ & 0.2174(3)  \\
                        	   && $\fZeroSinglet(0^{++})$ & 0.2007(18) \\
    $\fOneOctet(1^{++})$ & 0.3203(6)  	& $\fOneSinglet(1^{++})$ & 0.3364(14)  \\
    $\hOneOctet(1^{+-})$ & 0.3272(6) 	& $\hOneSinglet(1^{+-})$ & 0.3288(17)  \\    
\end{tabular}
\caption{Relevant stable hadron masses in temporal lattice units, $a_t m$.}
\label{tab:masses}
\end{table} 

The breaking of rotational symmetry on a cubic spatial lattice with a periodic cubic spatial boundary leads to irreducible representations, \emph{irreps}, of the reduced symmetry group which contain the subductions of states of definite angular momentum~\cite{Johnson:1982yq, Moore:2005dw, Thomas:2011rh}, as given in Table~\ref{tab:subductions}~\footnote{Note that Table~\ref{tab:subductions} corrects a typographical error in the corresponding table of Ref.~\cite{Johnson:2020ilc} which incorrectly indicated the presence of $2^{--}$ in $[111]\, A_1$. }.

\begin{table}[b]
\renewcommand{\arraystretch}{1.2}
\begin{tabular}{l|lll ll}
$[000]\, T_1^-$ & $1^{--}$ &          & $3^{--}$ \\
$[000]\, E^-$   &          & $2^{--}$ &          \\	
$[000]\, T_2^-$ &          & $2^{--}$ & $3^{--}$ \\
$[000]\, A_2^-$ &          &          & $3^{--}$ \\[1ex]
\hline\\[-2ex]
$[100]\, A_1$   & $1^{--}$ &          & $3^{--}$ & $\mathit{0^{+-}}$ & $\mathit{2^{+-}}$ \\
$[100]\, B_1$   &          & $2^{--}$ & $3^{--}$ &                   & $\mathit{2^{+-}}$ \\
$[100]\, B_2$   &          & $2^{--}$ & $3^{--}$ &                   & $\mathit{2^{+-}}$ \\[1ex]
\hline\\[-2ex]
$[111]\, A_1$   & $1^{--}$ &      & $\big(3^{--}\big)^2$ & $\mathit{0^{+-}}$ & $\mathit{2^{+-}}$ \\
\end{tabular}
\caption{Subductions of lowest $J^{PC}$ into cubic irreps with $C=-$; superscripts indicate multiple embeddings. }
\label{tab:subductions}
\end{table}

The possible meson-meson channels contributing at low energies to scattering in the relevant $J^{PC}$ are presented in Table~\ref{tab:partialwaves}, and the corresponding finite-volume non-interacting energies (computed as described in Ref.~\cite{Johnson:2020ilc}) are used to select the basis of operators by including all meson-meson operators~\cite{Thomas:2011rh} having a non-interacting energy below a certain cut-off (typically $a_t E_\mathsf{cm} \sim 0.46$). In addition to these, a large set of $q\bar{q}$-like fermion bilinear operators are included~\cite{Dudek:2010wm}.

\begin{table}
\renewcommand{\arraystretch}{1.3}
\begin{tabular}{r|l}
%
\multirow{ 2}{*}{$1^{--}$} & ${\color{jlab_red} \eta^\mathbf{8} \eta^\mathbf{8} \{ \onePone \} }, 
{\color{jlab_blue} \eta^\mathbf{8} \omega^\mathbf{8} \{ \threePone \} }, {\color{jlab_green} \eta^\mathbf{8} \omega^\mathbf{1} \{ \threePone \} }$ \\    
      & ${\color{jlab_gray} f_0^\mathbf{1} \omega^\mathbf{8} \{ \threeSone, \threeDone \},  \eta^\mathbf{1} \omega^\mathbf{8} \{ \threePone \}, \omega^\mathbf{8} \omega^\mathbf{8} \{ \threePone, \fivePone \}   } $ \\[1ex]
\multirow{ 2}{*}{$2^{--}$} &  ${\color{jlab_blue} \eta^\mathbf{8} \omega^\mathbf{8} \{ \threePtwo, \threeFtwo \}}, {\color{jlab_green}\eta^\mathbf{8} \omega^\mathbf{1} \{ \threePtwo, \threeFtwo \} }$ \\    
      & ${\color{jlab_gray} f_0^\mathbf{1} \omega^\mathbf{8} \{ \threeDtwo \},  \eta^\mathbf{1} \omega^\mathbf{8} \{ \threePtwo, \threeFtwo \}, \omega^\mathbf{8} \omega^\mathbf{8} \{ \fivePtwo \}   } $ \\[1ex]
\multirow{ 2}{*}{$3^{--}$} &  ${\color{jlab_red}\eta^\mathbf{8} \eta^\mathbf{8} \{ \oneFthree \} }, {\color{jlab_blue}\eta^\mathbf{8} \omega^\mathbf{8} \{ \threeFthree \} }, {\color{jlab_green}\eta^\mathbf{8} \omega^\mathbf{1} \{ \threeFthree \} }$ \\    
      & ${\color{jlab_gray} f_0^\mathbf{1} \omega^\mathbf{8} \{ \threeDthree \},  \eta^\mathbf{1} \omega^\mathbf{8} \{ \threeFthree \}, \omega^\mathbf{8} \omega^\mathbf{8} \{ \fivePthree, \oneFthree, \fiveFthree  \}   } $ \\[1ex]
\end{tabular}
\caption{Meson-meson scattering partial-waves for each $J^{P}$-- only waves with $\ell \leq 3$ shown. Channels in gray excluded from the coupled-channel amplitudes used to describe scattering in our selected energy region. }
\label{tab:partialwaves}
\end{table}

\begin{figure*}
\includegraphics[width=\textwidth]{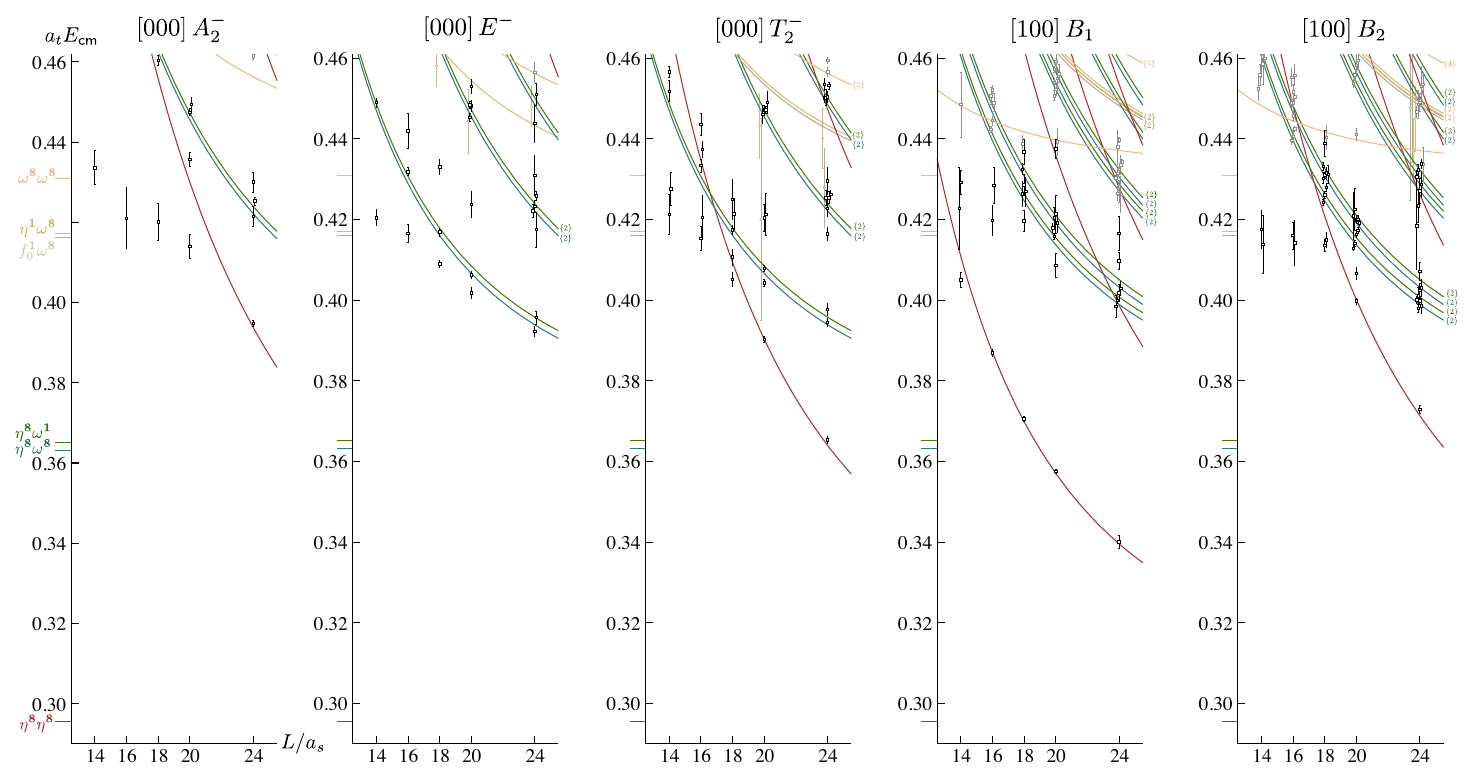}
\includegraphics[width=\textwidth]{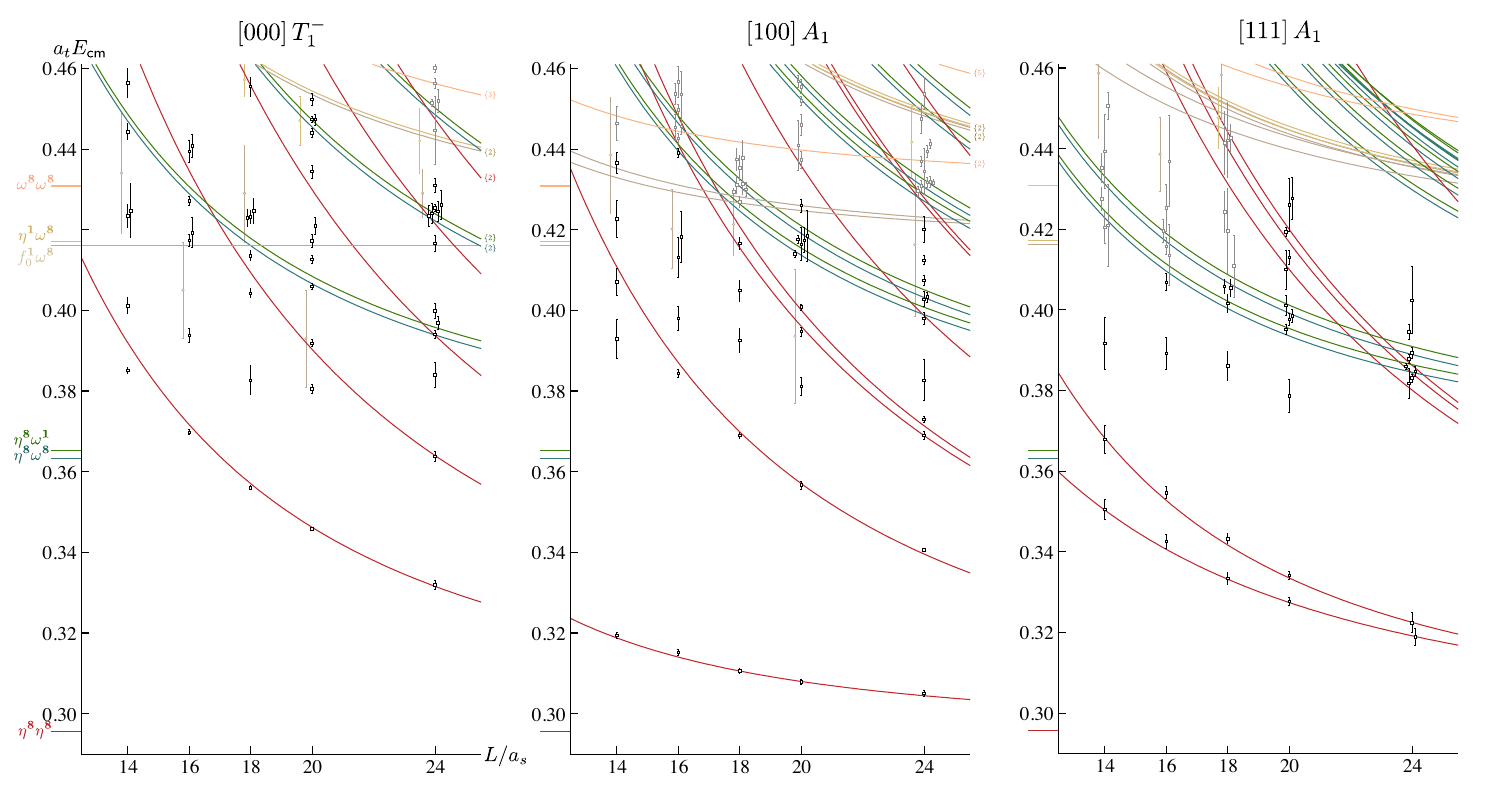}
\caption{Finite-volume spectra in eight irreps across five lattice volumes. Colored markers on vertical axis indicate kinematic thresholds for various meson-meson channels, while the colored curves show the corresponding non-interacting energies. Degeneracy of non-interacting levels is indicated by $\{n\}$ for $n>1$. Points with errorbars represent energies extracted from variational analysis of two-point correlator matrices, with some points displaced horizontally for clarity. Levels with dominant overlap onto decoupled channels $f_0^\mathbf{1} \omegaOctet$ (brown), $\etaSinglet \omegaOctet$ (sand) are indicated (only the lowest-lying such levels are shown) -- these, as well as all levels shown by gray points are not used to determine coupled-channel amplitudes.}
\label{fig:spec}
\end{figure*}

The computed matrices of correlation functions were analysed variationally as in previous \emph{hadspec} papers by solving generalized eigenvalue problems. A slight novelty in the current calculation is the use of a `model averaging' procedure which weights various eigenvalue timeslice fitting-windows and number of exponentials with a probability associated with a version of the Akaike Information Criterion~\cite{Jay:2020jkz, Radhakrishnan:2022ubg}. In practice this leads to modestly more conservative error estimates for some energy levels. The extracted discrete spectra across eight irreps are presented in Figure~\ref{fig:spec}~\footnote{In addition, spectra for irreps having leading contributions from the (exotic) positive parity partial-waves listed in Table~\ref{tab:subductions} were also computed, with no significant departures from non-interacting energies observed in the energy region of interest. Based on this we neglect scattering in these partial-waves in the remainder of the analysis. }.

We observe that we are able to determine large numbers of excited states, including many near-degenerate states in some energy regions having a high density of energy levels, making good use of the orthogonality properties of the generalized eigenvalue approach.
As well as the energies, the variational analysis also yields operator overlap factors for each state, and these indicate that some states, colored brown or sand in Figure~\ref{fig:spec}, overlap \emph{only} onto $f_0^\mathbf{1} \omegaOctet$ and $\etaSinglet \omegaOctet$ operators, suggesting that these channels are completely decoupled. We will exclude these levels from further analysis.
The operator overlaps show some evidence of coupling to the $\omegaOctet \omegaOctet$ channel, but only at the highest energies, above the energy where we anticipate resonances will lie. In order to avoid needing to consider $\omegaOctet \omegaOctet$ scattering, we will exclude all levels lying near to or above the lowest $\omegaOctet \omegaOctet$ non-interacting energy level in each irrep.

\medskip
The $[000]\, A_2^-$ spectrum shows a clear `extra' level, not lying close to any non-interacting energy, near $a_t E_\mathsf{cm} \sim 0.42$ on the smaller volumes, which is a likely indication of an isolated $3^{--}$ resonance.
In the $[000]\, E^-$ spectrum, as well as a consistently present level compatible with lying \emph{between} the lowest $\etaOctet \omegaOctet$, $\etaOctet \omegaSinglet$ non-interacting energies, we observe what might be an avoided-level crossing behavior around $a_t E_\mathsf{cm} \sim 0.42$. This would have a plausible explanation as a $2^{--}$ resonance coupled to $\etaOctet \omegaOctet$ and/or $\etaOctet \omegaSinglet$.
In $[000]\, T_2^-$, two `extra' levels are clearly visible in the $L/a_s = 14,20$ spectra, compatible with the $2^{--}$ and $3^{--}$ resonances proposed for the previous two irreps both being present here. Similar evidence for these two resonances is present in the $[100]\, B_{1,2}$ spectra.

The $[000]\,T_1^-$, $[100]\, A_1$ and $[111]\, A_1$ spectra into which $J^{PC}=1^{--}$ is subduced feature the stable $\omegaOctet$ as an isolated and essentially volume-independent level which is not plotted in Figure~\ref{fig:spec}. This bound-state lies well below the $\etaOctet \etaOctet$ threshold and is not expected to have a significant impact above threshold, and to the extent that it does, it should be reasonably modeled by a slowly varying `background' energy dependence in scattering amplitudes.

The $[000]\,T_1^-$ spectrum features several plausible avoided level crossings relative to $\etaOctet \etaOctet$ non-interacting energies near to $a_t E_\mathsf{cm} \sim 0.39$, which could be explained in terms of a low-lying $1^{--}$ resonance. Two `extra' levels visible near $a_t E_\mathsf{cm} \sim 0.42$ might have an interpretation as being due to the $3^{--}$ resonance inferred from the $[000]\, A_2^-$ irrep spectra, supplemented by a second $1^{--}$ resonance. Patterns in the $[100]\, A_1$ and $[111]\, A_1$ spectra would appear to be compatible with this proposed resonance content.

To go beyond these qualitative statements, we must analyse these spectra in terms of coupled-channel scattering amplitudes using the finite-volume quantization condition. We will make use of all levels presented as black squares in Figure~\ref{fig:spec}, amounting to 154 levels in the irreps having leading $2^{--}$ and $3^{--}$ subductions, and 119 levels in the irreps with leading $1^{--}$.

\section{Scattering Amplitudes}
  \label{sec:Amps}
 
The coupled-channel partial-wave $t$-matrix describing scattering in infinite volume is related to the discrete spectrum in a periodic $L\times L \times L$ box by the finite-volume quantization condition,
\begin{equation}
 0 = \det \Big[ \bm{1} + i \bm{\rho}(E_\mathsf{cm}) \, \bm{t}(E_\mathsf{cm}) \, \big( \bm{1} + i \bm{\mathcal{M}}(E_\mathsf{cm},L) \big) \Big] \, ,
\label{eq:luscher} 
\end{equation}
where the matrix $\bm{\mathcal{M}}$ contains known functions which encode the kinematics of the finite cubic volume. Ref.~\cite{Briceno:2017max} reviews the origin of this relation, and its contemporary application to energy levels obtained within lattice QCD calculations.

Eq.~\ref{eq:luscher} only has solutions for $t$-matrices which satisfy the constraint of multichannel unitarity, and parameterizations which achieve this are easily constructed using a $K$-matrix~\cite{Guo:2012hv},
\begin{equation}
\big[ \bm{t}^{-1}(s) \big]_{ij} = \frac{1}{(2 k)^{\ell_i}} \big[ \bm{K}^{-1}(s) \big]_{ij}  \frac{1}{(2 k)^{\ell_j}} + \bm{I}_{ij}(s) \, ,
\label{eq:tK}
\end{equation}
where the correct threshold behavior for a channel with orbital angular momentum $\ell$ is imposed. The matrix $\bm{K}(s)$ must be real and symmetric in the space of coupled-channels, while the diagonal matrix $\bm{I}(s)$ has an imaginary part given by the phase-space for each channel $i$, $-\rho_i(s)$. $\bm{I}(s)$ can optionally be given a real part, corresponding to dispersively improving $-\rho_i(s)$, in which case it is referred to as the ``Chew-Mandelstam phase-space''.

The likely presence of narrow resonances in the spectrum inspires the use of explicit poles in $\bm{K}(s)$, as they provide an efficient parameterization in such cases. As well as some number of poles, we may also consider including polynomial behavior in $s$, i.e.
\begin{equation}
K_{ij}(s) = \sum_p \frac{g^{(p)}_i \, g^{(p)}_j}{ (m^{(p)})^2 - s} + \gamma^{(0)}_{ij} + \gamma^{(1)}_{ij} s + \ldots
\end{equation}

For any given parameterization, and selection of parameter values, Eq.~\ref{eq:luscher} can be solved to yield the corresponding finite-volume spectrum for any desired $L$ value~\cite{Woss:2020cmp}, and this spectrum then compared to the computed lattice QCD spectra. Minimization of a correlated $\chi^2$ quantifying the comparison will provide best fit values for the amplitude parameters~\cite{Wilson:2014cna}.
The resulting $t$-matrix, with determined parameter values (and correlated uncertainties), can then be analysed for its pole content in the complex-energy plane,
\begin{equation}
  t_{ij}(s \sim s_0) \sim \frac{c_i\, c_j}{s_0 -s}\, ,
\label{eq:tpole}  
\end{equation}
where the pole location is typically interpreted in terms of the mass and width of a resonance, $\sqrt{s_0} = m_R \pm \tfrac{i}{2} \Gamma_R$, and the pole residue is factorized in terms of the resonance's complex-valued channel couplings, $c_i$.

This approach to determine scattering amplitudes from lattice QCD spectra is now well established, having been applied successfully to a range of coupled-channel scattering systems~\cite{Dudek:2014qha, Wilson:2014cna, Wilson:2015dqa, Moir:2016srx, Dudek:2016cru, Briceno:2017qmb, Woss:2019hse, Prelovsek:2020eiw, Lang:2022elg, BaryonScatteringBaSc:2023ori, PhysRevLett.132.241901, PhysRevD.109.114503}. 
Ultimately, its ability to constrain non-zero scattering amplitudes relies upon the departure of computed lattice QCD energy levels from non-interacting energy levels in a finite-volume. This leads to a particular challenge in the current calculation, where the near degeneracy of the stable $\omegaOctet$, $\omegaSinglet$ mesons in systems where there is the presence of both $\etaOctet \omegaOctet$ and $\etaOctet \omegaSinglet$ channels, leads to ambiguous scattering amplitude descriptions of the spectra. In order to illustrate the issue, we will explore it within a simple toy-model.

  \subsection{Separation of $\etaOctet\omegaOctet$, $\etaOctet\omegaSinglet$ channels.}
  \label{sec:toy_model}
 
The near degeneracy of the stable $\omegaOctet$ and $\omegaSinglet$ mesons ensures that $\etaOctet \omegaOctet$, $\etaOctet \omegaSinglet$ non-interacting levels always lie very close to each other in all volumes, as indicated by the blue and green curves in Figure~\ref{fig:spec}. One important effect of this is the limited ability of the finite-volume quantization condition, Eq.~\ref{eq:luscher}, to distinguish between scattering in these two channels. Here we will propose simple toy models of scattering in which the ratio of couplings of a resonance to these channels is a parameter, and explore the sensitivity of the finite-volume spectrum to this ratio.

\begin{figure*}
\includegraphics[width=\textwidth]{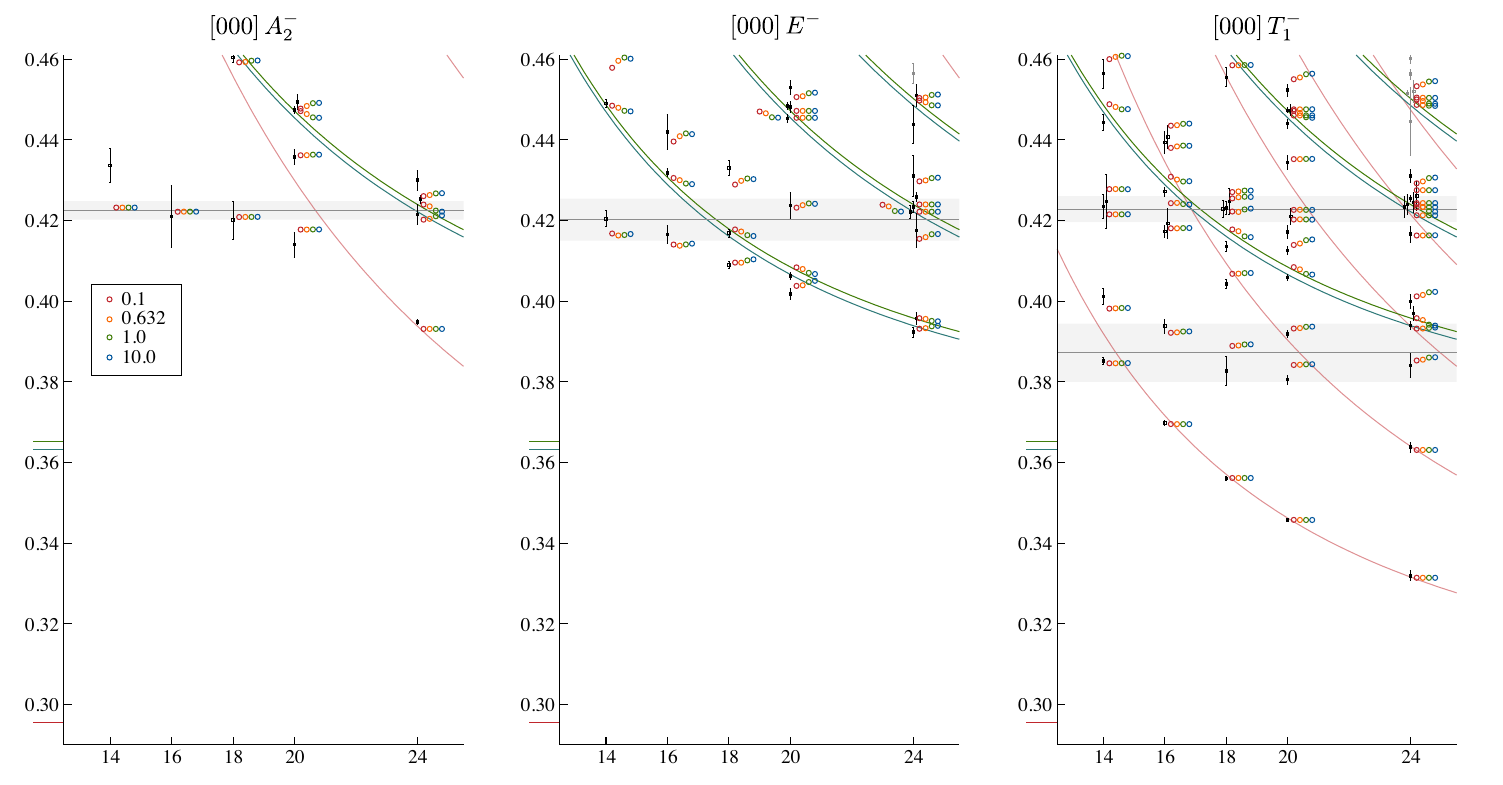}
\caption{Finite-volume spectra corresponding to simple toy models of resonance scattering having fixed coupling combination $ g_{\eta^\mathbf{8} \omega^\mathbf{8}}^{\, 2} + g_{\eta^\mathbf{8} \omega^\mathbf{1}}^{\,2} $, but varying ratio, $r = g_{\eta^\mathbf{8} \omega^\mathbf{1}}/ g_{\eta^\mathbf{8} \omega^\mathbf{8}} = 0.1, 0.632, 1.0, 10.0$, as described in the text. Gray bands indicate the toy model resonance masses and their widths.}
\label{fig:toy}
\end{figure*}

We begin by examining the $[000]\, A_2^-$ spectrum using a simple amplitude model of a single $3^{--}$ resonance coupled to $\etaOctet \etaOctet$, $\etaOctet \omegaOctet$ and $\etaOctet \omegaSinglet$. The  amplitude is constructed using a single $K$-matrix pole fixed in location at $a_t m=0.4227$, with a fixed $F$-wave coupling to $\eta^\mathbf{8} \eta^\mathbf{8}$ having a value inspired by fits reported later in this paper. Four values for the $K$-matrix \mbox{$F$-wave} pole couplings to $\eta^\mathbf{8} \omega^\mathbf{8}, \eta^\mathbf{8} \omega^\mathbf{1}$, are considered, where the ratio of the couplings, ${r \equiv g_{\eta^\mathbf{8} \omega^\mathbf{1}}/ g_{\eta^\mathbf{8} \omega^\mathbf{8}} }=0.1, 0.632, 1.0, 10.0$ is varied, 
while keeping $g_{\eta^\mathbf{8} \omega^\mathbf{8}}^{\, 2} + g_{\eta^\mathbf{8} \omega^\mathbf{1}}^{\,2} $ constant. The value ${r= \sqrt{2/5} \approx 0.632}$ is that proposed in Ref.~\cite{Johnson:2020ilc} to implement exactly the OZI-rule.

The corresponding finite-volume spectra in the $[000]\, A_2^-$ irrep, obtained by solving the finite-volume quantization condition for this amplitude, are presented in the leftmost panel of Figure~\ref{fig:toy} along with the lattice QCD computed spectra from Figure~\ref{fig:spec}. We emphasize that these colored points are not a fit to the computed data, but the amplitude parameter values are selected to make the comparison to the lattice QCD points meaningful. We can see that only those energy levels lying close to $\etaOctet \omegaOctet$, $\etaOctet \omegaSinglet$ non-interacting energies show any visible dependence upon $r$, and that even in those cases, the level of variation over two orders of magnitude change in $r$ is not significantly larger than the statistical uncertainty on the computed energy levels.

The middle panel of Figure~\ref{fig:toy}, for $[000]\, E^-$, shows a similar amplitude construction for a $2^{--}$ resonance coupled to $\etaOctet \omegaOctet$ and $\etaOctet \omegaSinglet$ with $P$-wave couplings whose ratio is varied as above, with the sum of squares again kept constant~\footnote{Modest $F$-wave couplings, required to get the correct number of energy levels in the spectrum, are kept fixed.}. Once again we see an insensitivity to the ratio of couplings.

The $[000]\, T_1^-$ irrep shown in the right panel of Figure~\ref{fig:toy} is described by an amplitude with \emph{two} $K$-matrix poles, each coupled to $\etaOctet \etaOctet$, $\etaOctet \omegaOctet$ and $\etaOctet \omegaSinglet$ in $P$-wave~\footnote{The $3^{--}$ amplitude of the leftmost panel is also included for this irrep.}. The same value of channel coupling ratio is used for the two poles, and again we see very little sensitivity in the spectrum to the value of $r$.

Moving-frame irrep spectra were also explored, and these showed no additional sensitivity to the value of $r$. In summary, it appears to be unlikely that fits to the lattice QCD computed spectra, having the current level of statistical precision, will be capable of constraining the ratio of couplings to the $\etaOctet \omegaOctet, \etaOctet \omegaSinglet$ channels. Explicit amplitude fits to our computed spectra presented later in this paper will confirm these expectations, finding that while the sum of squares of the couplings is rather precisely determined (as is the corresponding combination of \mbox{$t$-matrix} pole couplings), their ratio is poorly determined.

  \subsection{$2^{--}$, $3^{--}$ amplitudes.}
  \label{sec:J2J3}
 
We will proceed to describe 154 levels spread over five irreps and five volumes, as presented in the upper panels of Figure~\ref{fig:spec}, in terms of coupled-channel amplitudes describing $2^{--}$ and $3^{--}$ partial-waves. The set of channels needed are those presented in color in Table~\ref{tab:partialwaves}.

The simplest amplitudes that prove capable of describing the spectra feature just a single pole in each of the $2^{--}$ and $3^{--}$ \mbox{$K$-matrices}, and no additional polynomial, with use of the Chew-Mandelstam phase-space subtracted at the pole location. Given the discussion in the previous section, in both $J^{PC}$ cases we fix the ratio of $\etaOctet \omegaSinglet, \etaOctet \omegaOctet$ $K$-matrix pole couplings to $r=0.632$, in $P$-wave for $2^{--}$, and $F$-wave for $3^{--}$~\footnote{Both $F$-wave channel couplings in $2^{--}$ are allowed to freely float.}. The following parameter values give amplitudes that describe the 154 levels with a quite reasonable $\chi^2$, 
\begin{center}
\renewcommand{\arraystretch}{1.2}	
    \begin{tabular}{rll}
    $m(3^{--}) =$                         & $0.4227(9)\, a_t^{-1}$  &
    \multirow{3}{*}{ $\begin{bmatrix*}[r] 1 & -0.4 & 0.2 \\
                                            & 1    &  0.1 \\
                                            &      & 1    \\ 
                                            \end{bmatrix*}$ } \\
    $g_{\etaOctet\etaOctet; F}(3^{--})   = $                  &  $1.76(8)  \, a_t^2$   & \\
    $g_{\etaOctet\omegaOctet; F}(3^{--}) = $                  &  $-2.02(29)\, a_t^2$ & \\[2.2ex]
    $m(2^{--}) =$                         & $0.4209(10)\, a_t^{-1}$  &
    \multirow{4}{*}{ $\begin{bmatrix*}[r] 1 & 0.1  & -0.3 & 0.1 \\
                                            & 1    & 0.1  & 0.2 \\
                                            &      & 1    & -0.8 \\
                                            &      &      & 1 
                                            \end{bmatrix*}$ } \\
    $g_{\etaOctet\omegaOctet; P}(2^{--}) =   $            &  $0.369(14)$ & \\
    $g_{\etaOctet\omegaOctet; F}(2^{--}) =   $            &  $-1.3(6)  \, a_t^2$ & \\
    $g_{\etaOctet\omegaSinglet; F}(2^{--}) = $            &  $-1.9(7)  \, a_t^2$ & \\[1.3ex]
    \multicolumn{3}{l}{\quad\quad\quad\quad $\chi^2/N_{\text{dof}}=\frac{213.8}{154-7}=1.45$\,,}
    \end{tabular}
\end{center}\vspace{-0.9cm}
\begin{equation}\label{eq:J2J3refamp}\end{equation}
where the only large parameter (anti-)correlation is one we could have anticipated, between the floating $F$-wave $\etaOctet \omegaOctet$ and $\etaOctet \omegaSinglet$ couplings~\footnote{Not explicitly shown is the fact that no large correlations were found between the $2^{--}$ and $3^{--}$ amplitude parameters.}.

\begin{figure}
\includegraphics[width=\columnwidth]{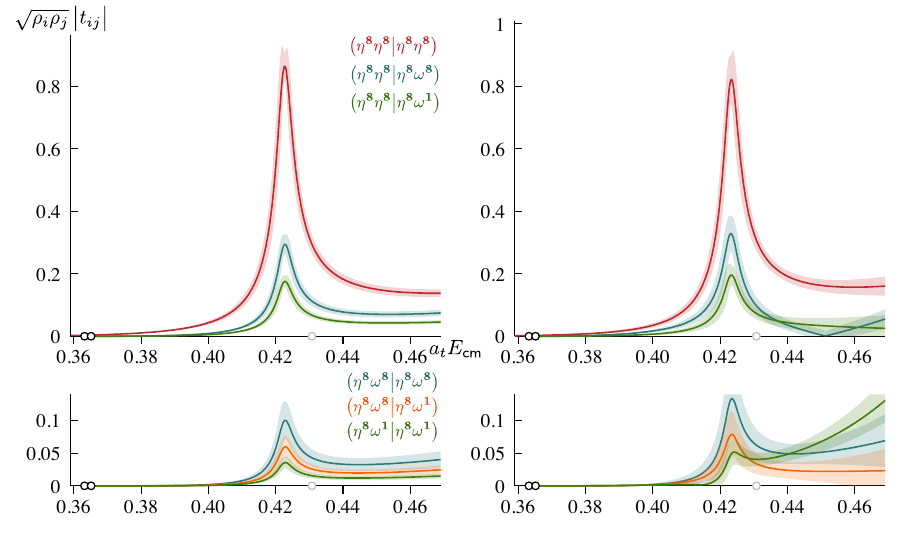}
\caption{$3^{--}$ coupled-channel scattering amplitudes with coupling ratio $g_{\etaOctet \omegaSinglet} / g_{\etaOctet \omegaOctet}$ fixed to value 0.632. Left panel: single $K$-matrix pole (see Eq.~\ref{eq:J2J3refamp}). Right panel: $K$-matrix pole plus constants. }
\label{fig:3m_amps}
\end{figure}

\begin{figure}
\includegraphics[width=\columnwidth]{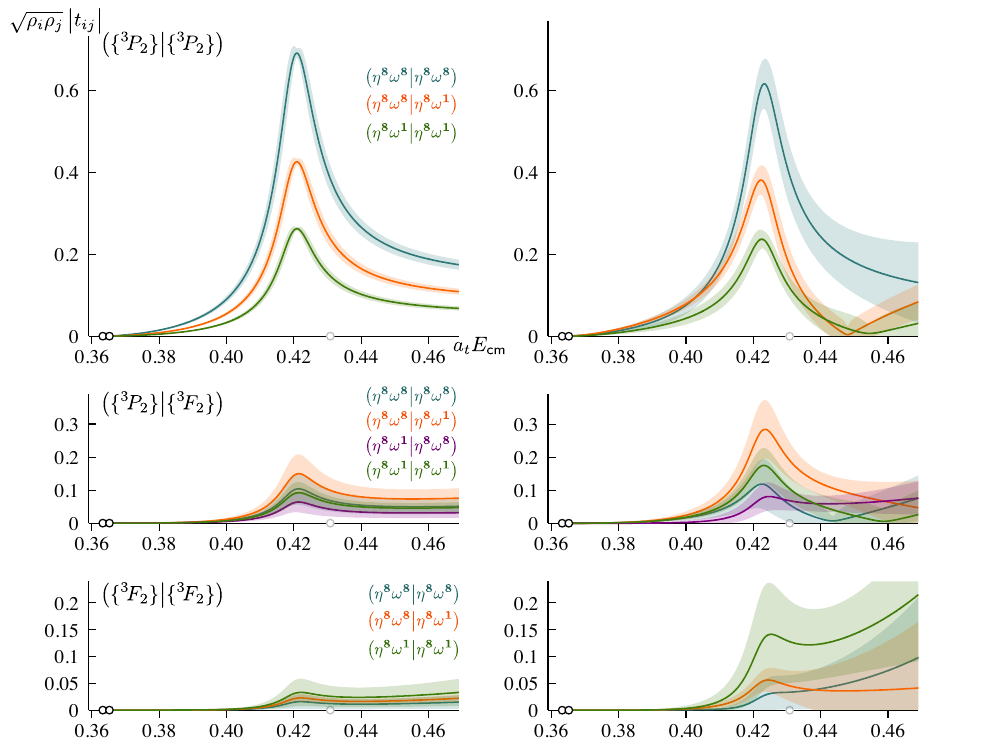}
\caption{$2^{--}$ coupled-channel scattering amplitudes with \mbox{$P$-wave} coupling ratio $g_{\etaOctet \omegaSinglet;P} / g_{\etaOctet \omegaOctet;P}$ fixed to value 0.632. Left panel: single $K$-matrix pole (see Eq.~\ref{eq:J2J3refamp}). Right panel: $K$-matrix pole plus constants. }
\label{fig:2m_amps}
\end{figure}

The obtained amplitudes are shown\footnote{In order to be able to better resolve weaker amplitudes, we are choosing to plot $\rho |t|$ in this paper, rather than $\rho^2 |t|^2$ as we have in previous \emph{hadspec} papers.} in the leftmost panels of Figures~\ref{fig:3m_amps}, \ref{fig:2m_amps}, where we see clear narrow peaks in both $3^{--}$ and $2^{--}$. While the relative height of the peaks in $t$-matrix elements including either $\etaOctet \omegaSinglet$ or $\etaOctet \omegaOctet$ will be sensitive to the fixed value of $r$, and will change if we change the value of $r$, these plots indicate the level of statistical precision that can be obtained from the spectra presented in Figure~\ref{fig:spec}.

The $3^{--}$ amplitude of Eq.~\ref{eq:J2J3refamp} is found to house a \mbox{$t$-matrix} pole on the proximal sheet above $\etaOctet \etaOctet, \etaOctet \omegaOctet, \etaOctet \omegaSinglet$ thresholds, located at
\begin{equation*}
a_t \sqrt{s_0}    = 0.4225(9) \pm \tfrac{i}{2}\, 0.0046(4)  \, ,
\end{equation*}
with channel couplings,
\begin{align*}
a_t c_{\etaOctet \etaOctet\{\oneFthree\}} &= 0.049(2)\, e^{\mp i \pi 0.023(2) }  \, ,    \\[1.2ex]
a_t c_{\etaOctet \omegaOctet\{\threeFthree\}}   &= 0.020(3)\, e^{\pm i \pi 0.967(3) } \, ,      \\
a_t c_{\etaOctet \omegaSinglet\{\threeFthree\}} &= 0.012(2)\, e^{\pm i \pi 0.966(3) }  \, ,   
\end{align*}
which we observe to be close to being real-valued. The ratio of the magnitudes of $\etaOctet \omegaSinglet$, $\etaOctet \omegaOctet$ pole couplings,
\begin{equation*}
\left| \frac{c_{\etaOctet \omegaSinglet\{\threeFthree\}}}{c_{\etaOctet \omegaOctet\{\threeFthree\}}} \right| = 0.6012(5) \, ,
\end{equation*}
clearly inherits a value close to the fixed $r=0.632$ of the $K$-matrix pole couplings.

\pagebreak

The $2^{--}$ amplitude of Eq.~\ref{eq:J2J3refamp} is found to house a \mbox{$t$-matrix}  pole on the proximal sheet above both $\etaOctet \omegaOctet, \etaOctet \omegaSinglet$ thresholds, located at
\begin{equation*}
a_t \sqrt{s_0}    = 0.4202(10) \pm \tfrac{i}{2}\, 0.0104(8) \, ,
\end{equation*}
with channel couplings,
\begin{align*}
a_t c_{\etaOctet \omegaOctet\{\threePtwo\}}   &= 0.078(3)\, e^{\mp i \pi 0.04(3) }   \, ,   \\
a_t c_{\etaOctet \omegaSinglet\{\threePtwo\}} &= 0.048(2)\, e^{\mp i \pi 0.04(3) }   \, ,   \\[1.2ex]
a_t c_{\etaOctet \omegaOctet\{\threeFtwo\}}   &= 0.012(6)\, e^{\pm i \pi 0.930(5) }   \, ,   \\
a_t c_{\etaOctet \omegaSinglet\{\threeFtwo\}} &= 0.017(6)\, e^{\pm i \pi 0.928(5) }     \, ,
\end{align*}
which are all observed to be close to real-valued. As we might expect, the $P$-wave couplings are significantly larger than the $F$-wave couplings, which are rather imprecisely determined.
The ratio of $P$-wave couplings,
\begin{align*}
\left| \frac{c_{\etaOctet \omegaSinglet\{\threePtwo\}}}{c_{\etaOctet \omegaOctet\{\threePtwo\}}}  \right|
&= 0.6212(2)  \, , 
\end{align*}
again inherits the fixed $r=0.632$ of the $K$-matrix pole coupling, while the $F$-wave coupling ratio,
\begin{align*}
\left| \frac{c_{\etaOctet \omegaSinglet\{\threeFtwo\}}}{c_{\etaOctet \omegaOctet\{\threeFtwo\}}}  \right|
&= 1.5 \pm 1.1 \, ,
\end{align*}
is rather imprecise.

\bigskip

Use of a pure single-pole form for the $K$-matrix forces a factorization of the $t$-matrix for \emph{all} real energy values, rendering it rank one, and this is not physically well motivated in general. A straightforward way to loosen this assumption is to add to the $K$-matrix pole a symmetric matrix of real constants. If the spectrum is described allowing such constants in the $3^{--}$ $K$-matrix, a $\chi^2/N_\mathrm{dof} = 193.2/(154-13) = 1.37$ is obtained, which is seen to be slightly lower than that in Eq.~\ref{eq:J2J3refamp}. The corresponding amplitude is shown in the right panel of Figure~\ref{fig:3m_amps}, where we observe that the strong peak feature is qualitatively unchanged.

An analogous addition of a matrix of constants to the $2^{--}$ $K$-matrix, yields a ${\chi^2/N_\mathrm{dof} = 177.4/(154-17) = 1.29}$, which is again a modest improvement over Eq.~\ref{eq:J2J3refamp}. The amplitude, presented in the right panel of Figure~\ref{fig:2m_amps}, shows the same peak structure, with some mild adjustment of the relative strength of $F$-wave versus $P$-wave.

\begin{figure}
\includegraphics[width=\columnwidth]{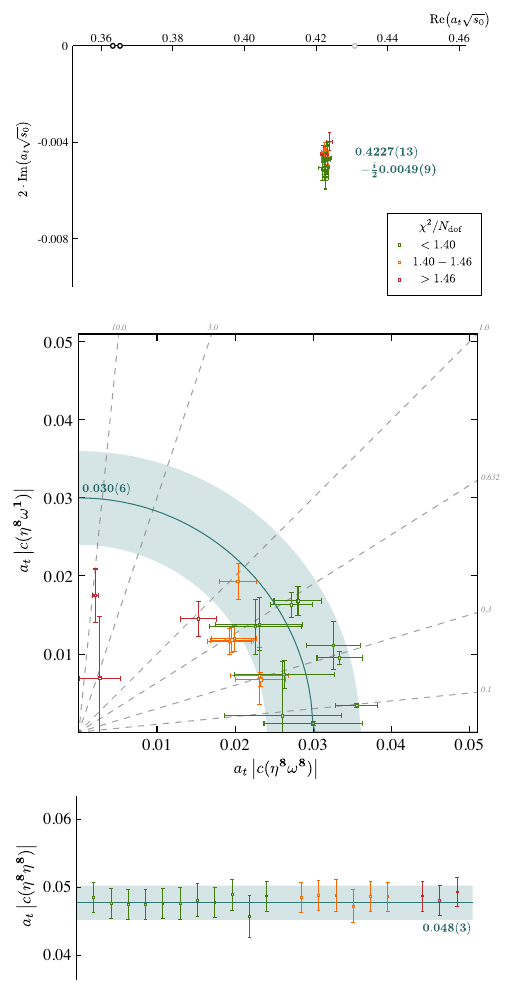}
\caption{Resonance pole in $3^{--}$ $t$-matrix over a range of parameterizations, including variation of fixed value of $r$. Top panel: pole location in lower half-plane of proximal sheet. Numerical value quoted accounts for parameterization variation. Middle panel: Magnitude of pole couplings to $\etaOctet \omegaSinglet$, $\etaOctet \omegaOctet$. Blue quarter-circle shows a conservative estimate of $\left(|c_{\etaOctet \omegaSinglet}|^2 + |c_{\etaOctet \omegaOctet}|^2 \right)^{1/2}$. Bottom panel: Magnitude of pole coupling to $\etaOctet \etaOctet$.  }
\label{fig:3m_pole}
\end{figure}

\begin{figure}
\includegraphics[width=\columnwidth]{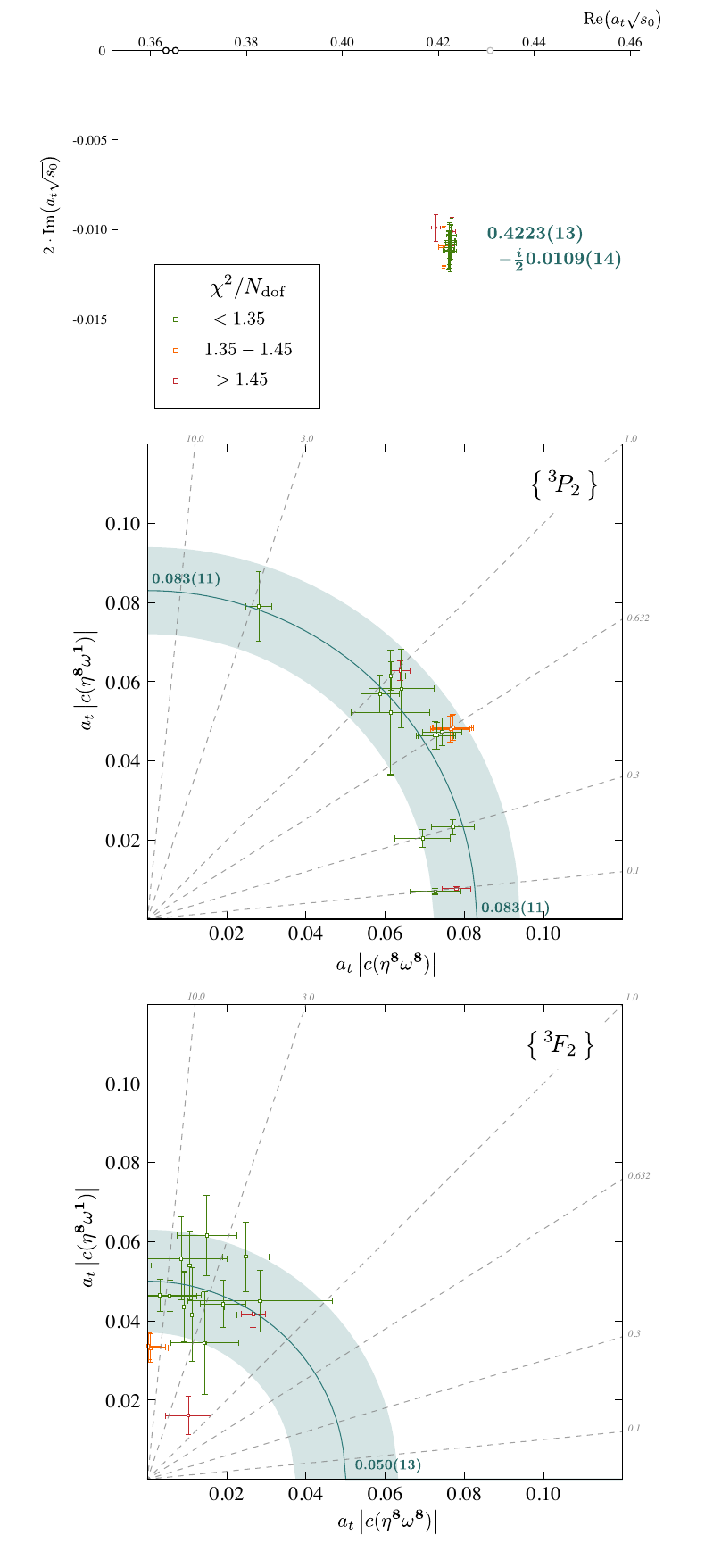}
\caption{Resonance pole in $2^{--}$ $t$-matrix over a range of parameterizations, including variation of fixed value of the ratio of $K$-matrix $P$-wave pole couplings to $\etaOctet \omegaSinglet$, $\etaOctet \omegaOctet$. Top panel: pole location in lower half-plane of proximal sheet. Numerical value quoted accounts for parameterization variation. Middle panel: Magnitude of pole couplings to $\etaOctet \omegaSinglet$, $\etaOctet \omegaOctet$ in $P$-wave. Blue quarter-circle shows a conservative estimate of $\left(|c_{\etaOctet \omegaSinglet}|^2 + |c_{\etaOctet \omegaOctet}|^2 \right)^{1/2}$. Bottom panel: Magnitude of pole couplings to $\etaOctet \omegaSinglet$, $\etaOctet \omegaOctet$ in $F$-wave.}
\label{fig:2m_pole}
\end{figure}

\bigskip

In order to assess sensitivity to the particular amplitude parameterization form, and to the fixed value of $r$, we attempted to describe the finite-volume spectra using a range of choices, including allowing constants to be added to the $K$-matrix poles, replacing such constants with a matrix proportional to $s$, and using the simple phase-space, rather than the Chew-Mandelstam form. For the $3^{--}$ case, the results of this variation for the resonance pole and its couplings are presented in Figure~\ref{fig:3m_pole}. The points are color-coded according to the $\chi^2/N_\mathrm{dof}$ with which they describe the spectrum. The top panel shows that the lattice QCD spectrum requires there to be a $3^{--}$ resonance whose pole position is very well determined, irrespective of the details of the parameterization, or the value of $r$ selected. The middle panel shows, as anticipated, that while the couplings to $\etaOctet \omegaSinglet$ and $\etaOctet \omegaOctet$ are not individually well determined, the value of the combination ${ \big( |c_{\etaOctet \omegaSinglet}|^2 + |c_{\etaOctet \omegaOctet}|^2 \big)^{1/2}}$ (indicated by the blue quarter-circle) \emph{can} be established. The plot indicates that, on the basis of $\chi^2$, there is perhaps a slight preference for a ratio of couplings less than unity in the $3^{--}$ case. The bottom panel shows that the pole coupling to $\etaOctet \etaOctet$ is rather well determined, regardless of the details of the parameterization, or the fixed value of $r$.

Variations of the $2^{--}$ amplitude lead to resonance pole variations shown in Figure~\ref{fig:2m_pole}, where again we observe that the lattice QCD spectrum demands there is a pole on the proximal sheet at a location which is quite precisely determined. The $P$-wave couplings to $\etaOctet \omegaSinglet$ and $\etaOctet \omegaOctet$ are again not individually determined, but the sum of their squares is. The ratio of $F$-wave $K$-matrix pole couplings is not fixed in our procedure, and we see that description of the spectrum suggests a ratio of such couplings that may be larger than unity, albeit with the couplings being not very precisely determined.

  \subsection{$1^{--}$ amplitudes}
  \label{sec:J1}
 
As discussed in Section~\ref{sec:Spectrum}, the spectra in the lower panels of Figure~\ref{fig:spec} suggest the presence of \emph{two} $1^{--}$ resonances. We will begin an investigation of this by presenting, as an example,  one successful scattering amplitude description of all 119 energy levels across the $[000]\, T_1^-$, $[100]\, A_1$, and $[111]\, A_1$ irreps.

Since the $3^{--}$ partial wave also subduces into these irreps, we must include it in the finite-volume quantization condition -- we choose to fix the $3^{--}$ amplitude to one obtained in the previous section\footnote{A single pole $K$-matrix with $r=0.632$ -- sensitivity to this particular amplitude choice was found to be weak.}, and vary only parameters in the $1^{--}$ amplitude parameterization.

The $1^{--}$ amplitude we will use as illustration features two $K$-matrix poles, where the pole coupling ratios, $r^{(0)} = g^{(0)}_{\etaOctet\omegaSinglet} / g^{(0)}_{\etaOctet\omegaOctet}$ and $r^{(1)} = g^{(1)}_{\etaOctet\omegaSinglet} / g^{(1)}_{\etaOctet\omegaOctet}$ are each fixed to value 0.632. These poles are supplemented with a complete symmetric matrix of unfixed constants, and use of the Chew-Mandelstam phase-space subtracted at $s=(m^{(0)})^2$, the location of the lower-mass $K$-matrix pole. A best-fit description of the finite-volume spectra is obtained with parameter values,
\begin{center}
{
\scriptsize
\renewcommand{\arraystretch}{1.6}	
    \begin{tabular}{rll}
    $m^{(0)} =$                         & $0.3891(13)\, a_t^{-1}$  &
    \multirow{3}{*}{ $\begin{bmatrix*}[r] 1 & 0.2 & -0.4 \\
                                            & 1    &  0.2 \\
                                            &      & 1    \\ 
                                            \end{bmatrix*}$ } \\
    $g^{(0)}_{\etaOctet\etaOctet}   = $                  &  $0.227(13)$   & \\
    $g^{(0)}_{\etaOctet\omegaOctet} = $                  &  $0.882(58)$ & \\[2.2ex]
    $m^{(1)} =$                         & $0.4198(11)\, a_t^{-1}$  &
    \multirow{3}{*}{ $\begin{bmatrix*}[r] 1 & -0.2  & -0.1  \\
                                            & 1    & 0.1   \\
                                            &      & 1   \\
                                            \end{bmatrix*}$ } \\
    $g^{(1)}_{\etaOctet\etaOctet} =   $            &  $-0.194(20)$ & \\
    $g^{(1)}_{\etaOctet\omegaOctet} =   $            &  $0.173(38)$ & \\[2.2ex]
    $\gamma(\etaOctet\etaOctet| \etaOctet \etaOctet) = $ & $(-0.49 \pm 0.47)\, a_t^2$ &
    \multirow{6}{*}{ $\begin{bmatrix*}[r] 1 & -0.1  & 0.0  & 0.0  & -0.1  & -0.1  \\
                                            & 1     & 0.1  & 0.4  & 0.4   & -0.3  \\
                                            &       & 1    & 0.1  & 0.2   & 0.3   \\
                                            &       &      & 1    & 0.9   & -0.4  \\
                                            &       &      &      & 1     & -0.1  \\
                                            &       &      &      &       & 1 \\
                                            \end{bmatrix*}$ } \\
    $\gamma(\etaOctet\etaOctet| \etaOctet \omegaSinglet) = $          & $(2.2 \pm 1.8)\, a_t^2$ \\
    $\gamma(\etaOctet\etaOctet| \etaOctet \omegaOctet) = $            & $(-1.7 \pm 1.3)\, a_t^2$ \\
    $\gamma(\etaOctet\omegaSinglet| \etaOctet \omegaSinglet) = $      & $(9.1 \pm 3.5)\, a_t^2$ \\	
    $\gamma(\etaOctet\omegaSinglet| \etaOctet \omegaOctet) = $        & $(3.9 \pm 2.4)\, a_t^2$ \\
    $\gamma(\etaOctet\omegaOctet| \etaOctet \omegaOctet) = $          & $(9.2 \pm 3.8)\, a_t^2$ \\
    \multicolumn{3}{l}{\quad\quad\quad\quad $\chi^2/N_{\text{dof}}=\frac{175.4}{119-12}=1.64$\,,}
    \end{tabular}
}    
\end{center}\vspace{-0.9cm}
\begin{equation}\label{eq:J1refamp}\end{equation}
where the only significant parameter correlations not shown above are between $g^{(0)}_{\etaOctet\omegaOctet}$ and  $\gamma(\etaOctet\omegaSinglet| \etaOctet \omegaOctet)$ and $\gamma(\etaOctet\omegaOctet| \etaOctet \omegaOctet)$. 

\begin{figure}
\includegraphics[width=\columnwidth]{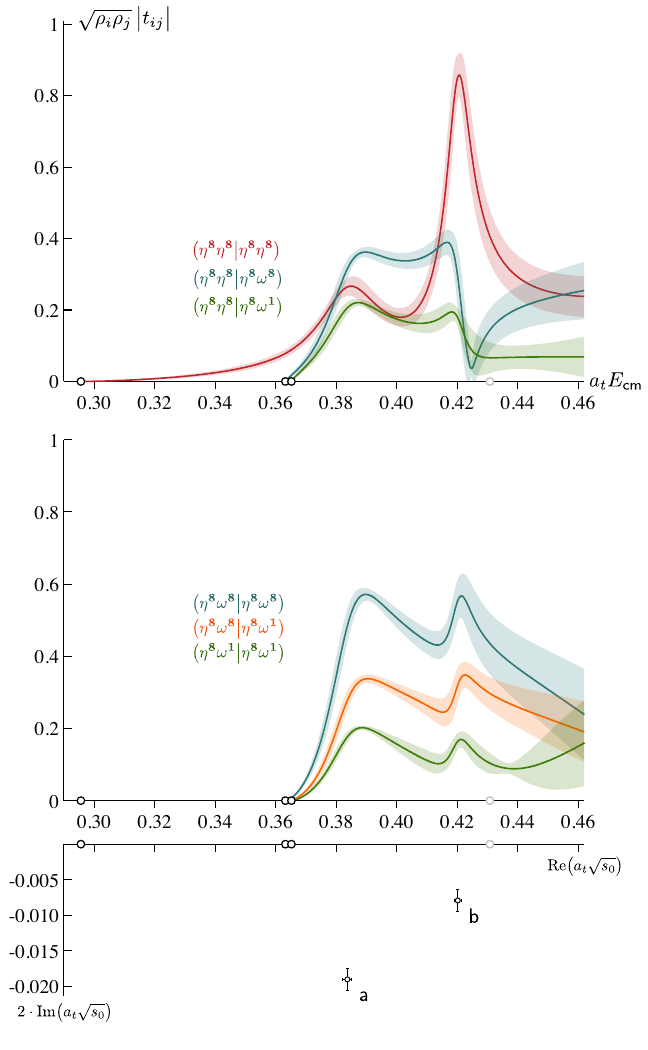}
\caption{$1^{--}$ coupled-channel scattering amplitudes of Eq.~\ref{eq:J1refamp}, together with $t$-matrix pole content on the proximal sheet above $\etaOctet\etaOctet, \etaOctet\omegaOctet, \etaOctet \omegaSinglet$ thresholds.  }
\label{fig:1m_amps}
\end{figure}

The $t$-matrix elements corresponding to the above description are shown in Figure~\ref{fig:1m_amps}, along with the $t$-matrix poles on the proximal sheet above $\etaOctet\etaOctet, \etaOctet\omegaOctet, \etaOctet \omegaSinglet$ thresholds.
As was also found to be the case in the $SU(3)$ flavor \emph{singlet} case in Ref.~\cite{Johnson:2020ilc}, the lighter resonance has a larger total width than the narrow heavier resonance. Notably, while the singlet case is a single-channel process, and hence tightly constrained by elastic unitarity, such that interference between the two resonances was forced to manifest as a sharp dip, the octet case is a system of coupled channels, where overlapping resonances can present in one of several ways. In fact we see that depending upon the element of the $t$-matrix considered, the two resonances can appear as two peaks, as a peak and a dip (in $t(\etaOctet \etaOctet|\etaOctet \omegaOctet)$), or as a peak and a `shoulder' (in $t(\etaOctet \etaOctet|\etaOctet \omegaSinglet)$).

\begin{figure}
\includegraphics[width=\columnwidth]{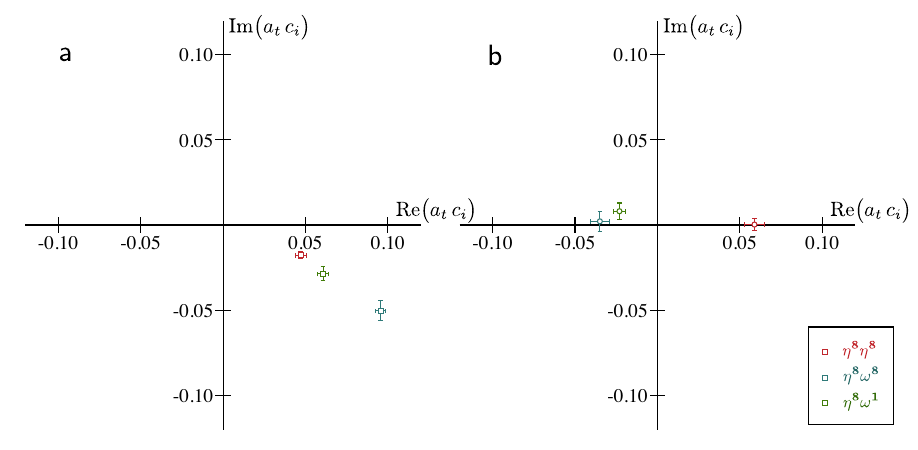}
\caption{Complex-valued $t$-matrix pole couplings for $1^{--}$ scattering amplitudes of Eq.~\ref{eq:J1refamp}, with poles labelled as in Figure~\ref{fig:1m_amps}. }
\label{fig:1m_pole_couplings_complex}
\end{figure}

In Figure~\ref{fig:1m_pole_couplings_complex} we present factorized residue couplings at the two $t$-matrix poles, where the heavier, narrower pole ($\mathsf{b}$) is observed to have couplings which are nearly real, while the lighter, broader pole ($\mathsf{a}$) has couplings which have a nonzero phase. The two poles are found to have comparable magnitude of coupling to $\etaOctet \etaOctet$, while $\mathsf{a}$ has significantly larger couplings to $\etaOctet \omegaOctet, \etaOctet \omegaSinglet$ than $\mathsf{b}$. We remind the reader than the relative size of couplings to $\etaOctet \omegaOctet$ and $\etaOctet \omegaSinglet$ for each resonance, and hence the magnitude of different $t$-matrix elements in Figure~\ref{fig:1m_amps}, is at this stage essentially determined by our choices of fixed $r^{(0)}, r^{(1)}$. Nevertheless, it should be clear that the finite-volume spectrum computed in lattice QCD leads to statistically quite precise amplitudes within this model.

\begin{figure*}
\includegraphics[width=0.92\textwidth]{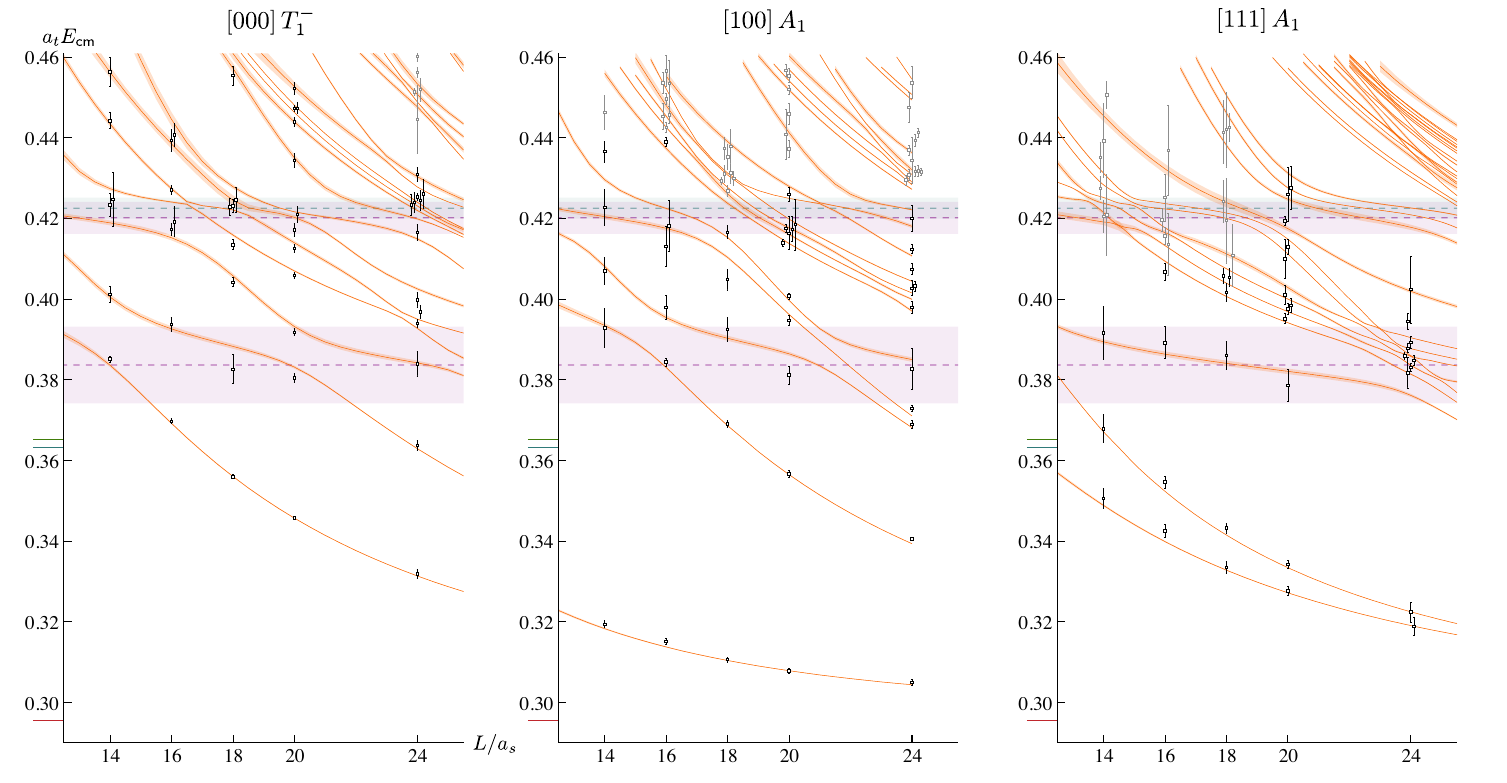}
\caption{Finite-volume spectra corresponding to amplitude in Eq.~\ref{eq:J1refamp} (orange curves) compared to computed lattice QCD spectra (black and gray points). Purple and blue horizontal bands indicate the masses and widths of the $1^{--}$ and $3^{--}$ resonances present in the amplitudes.}
\label{fig:1m_orange_curves}
\end{figure*}

The quality of the description of the finite-volume spectra can be seen in Figure~\ref{fig:1m_orange_curves}, where the orange curves correspond to the spectrum obtained by solving the finite-volume quantization condition for the amplitude of Eq.~\ref{eq:J1refamp} (and the fixed $3^{--}$ amplitude). This amplitude is observed to give a faithful description of the lattice QCD computed spectrum used to constrain it, and in fact appears to quite well reproduce the pattern of many levels (in gray) that were not included in the $\chi^2$ minimization\footnote{Our ability to make use of densely-packed energy levels in explicit fits is somewhat limited by our current implementation of algorithms to find solutions of the quantization condition and match them to computed lattice QCD energy levels. This is done many hundreds or thousands of times during a $\chi^2$ minimization, and hence must be extremely reliable. Further improvements beyond those described in Ref.~\cite{Woss:2020cmp} are under current investigation.}.

\bigskip

In Section~\ref{sec:toy_model} we presented an argument which indicates that we do not expect the ratios of couplings to $\etaOctet \omegaOctet$ and $\etaOctet \omegaSinglet$ to be well determined using the constraint supplied by our finite-volume spectra, but nevertheless, we can \emph{attempt} to describe the finite-volume spectra allowing the couplings ratios to float.
As an illustration, consider a fit to the 46 energy levels in the $[000]\, T_1^-$ irrep using an amplitude built from a $1^{--}$ $K$-matrix featuring just two poles and no constant terms, with $r^{(1)}$ (the $\etaOctet \omegaSinglet, \etaOctet \omegaOctet$ coupling ratio for the higher mass pole) fixed to 0.632, but $r^{(0)}$ allowed to float in the fit. Such a fit yields a quite reasonable $\chi^2/N_\mathrm{dof} = 62.5/(46-7) = 1.60$, with parameter values,
\begin{center}
{
\renewcommand{\arraystretch}{1.6}	
    \begin{tabular}{rll}
    $m^{(0)} =$                         & $0.3890(13)\, a_t^{-1}$  &
    \multirow{4}{*}{ $\begin{bmatrix*}[r] 1 & 0.2  & 0.1  & -0.1 \\
                                            & 1    & 0.1  & 0.1 \\
                                            &      & 1    & {\color{jlab_red}\bm{-1}}\\
                                            &      &      & 1 
                                            \end{bmatrix*}$ } \\
    $g^{(0)}_{\etaOctet\etaOctet}   = $                  &  $0.181(12)$   & \\
    $g^{(0)}_{\etaOctet\omegaOctet} = $                  &  $0.227(161)$ & \\
    $r^{(0)}                        = $                  &  $3.17 \pm 2.43$ & \\[2.2ex]
    $m^{(1)} =$                         & $0.4230(16)\, a_t^{-1}$  &
    \multirow{3}{*}{ $\begin{bmatrix*}[r] 1 & -0.1  & 0.0  \\
                                            & 1    & 0.4   \\
                                            &      & 1   \\
                                            \end{bmatrix*}$ } \\
    $g^{(1)}_{\etaOctet\etaOctet} =   $            &  $-0.207(25)$ & \\
    $g^{(1)}_{\etaOctet\omegaOctet} =   $            &  $0.120(58)$ & \\[2.2ex]
    \end{tabular} 
}   
\end{center}\vspace{-0.9cm}
\begin{equation}\label{eq:J1_float_r} \end{equation}
where none of the (not shown) parameter correlations between the two poles is larger than 0.3.
We observe that while most parameters are determined with good statistical precision, there are very large errors on $g^{(0)}_{\etaOctet\omegaOctet}$ and $r^{(0)}$, but that these parameters are 100\% anti-correlated, such that if this anti-correlation is accounted for,
\begin{align*}
&\left( \big(g^{(0)}_{\etaOctet \omegaOctet} \big)^2 + \big(g^{(0)}_{\etaOctet \omegaSinglet} \big)^2 \right)^{1/2} \\
&\quad\quad= \left| g^{(0)}_{\etaOctet \omegaOctet} \right| \left( 1 + \big(r^{(0)} \big)^2 \right)^{1/2} = 0.761(20) \, ,
\end{align*}
which is rather precisely determined. 
This anti-correlation between $r$ and $g_{\etaOctet \omegaOctet}$ proves to be a general feature of attempts to allow the ratio of couplings to float for either the lower or higher pole. This observation, along with the fairly precise determination of the sum of squares of the $\etaOctet \omegaOctet, \etaOctet \omegaSinglet$ couplings is exactly what was indicated by the toy model analysis presented in Section~\ref{sec:toy_model}. As such, as was the case for the $2^{--}, 3^{--}$ presented in Section~\ref{sec:J2J3}, we will proceed by considering amplitude variations in which the ratios $r^{(0)}, r^{(1)}$ are fixed to various values, allowing other parameters to float.

\begin{figure*}
\includegraphics[width=\textwidth]{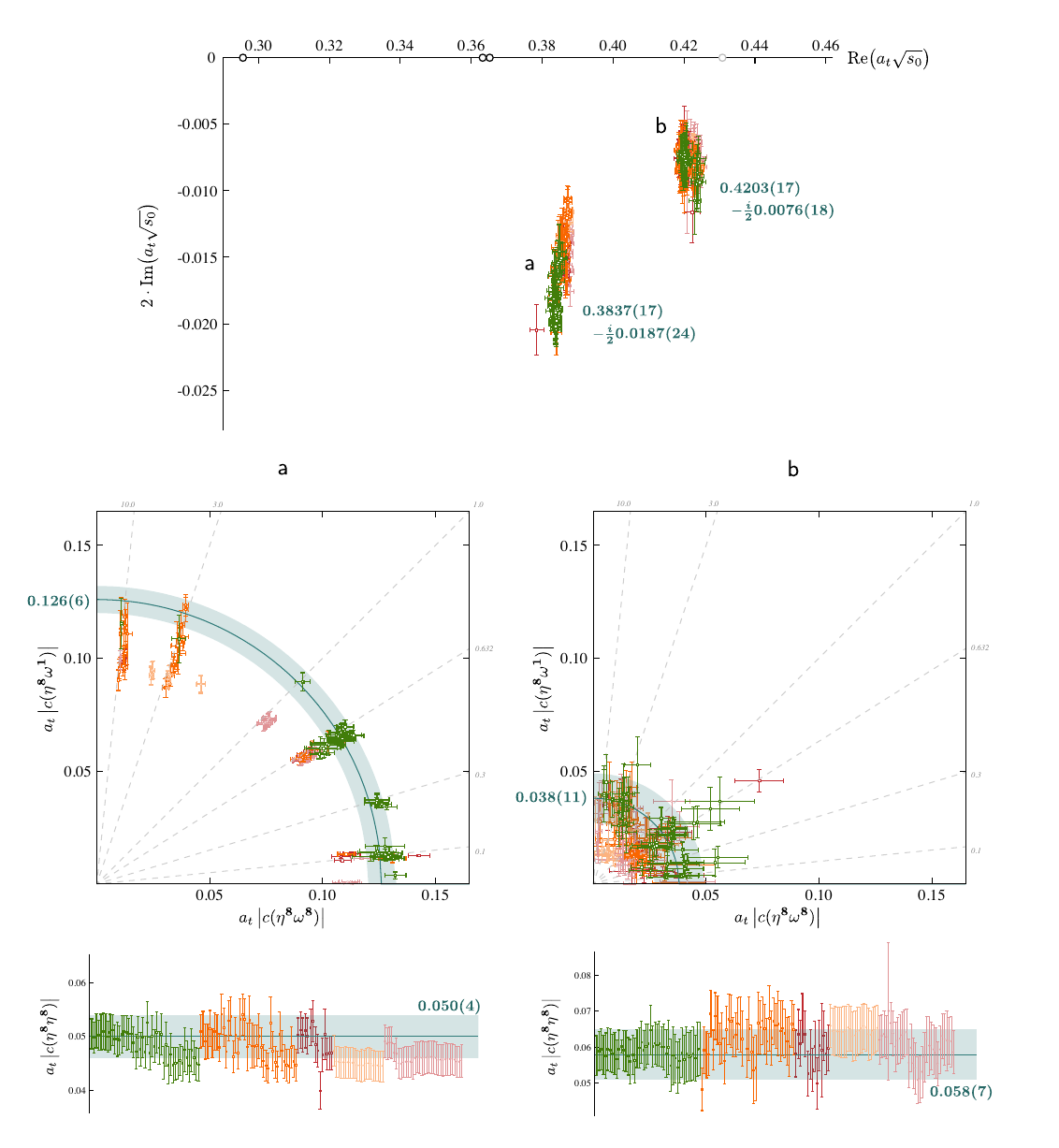}
\caption{Resonance poles in $1^{--}$ $t$-matrix over a range of parameterizations on the proximal sheet above $\etaOctet \etaOctet, \etaOctet \omegaOctet, \etaOctet \omegaSinglet$ thresholds. Top panel: pole location in lower half-plane of proximal sheet. Numerical value quoted accounts for parameterization variation. Middle panels: Magnitude of $P$-wave pole couplings to $\etaOctet \omegaSinglet$, $\etaOctet \omegaOctet$. Blue quarter-circles show a conservative estimate of $\left(|c_{\etaOctet \omegaSinglet}|^2 + |c_{\etaOctet \omegaOctet}|^2 \right)^{1/2}$. Bottom panels: Magnitude of pole coupling to $\etaOctet \etaOctet$.
Green points have low $\chi^2$, orange points somewhat larger $\chi^2$ and red points larger still. Thinnest lines correspond to fitting only $[000]\, T_1^-$, medium thickness to $[000]\, T_1^-, [100]\, A_1$, and thickest to all 119 levels in $[000]\, T_1^-, [100]\, A_1$ and $[111]\, A_1$. Lighter color points are for an amplitude having two $K$-matrix poles but no constants. }
\label{fig:1m_poles}
\end{figure*}

\bigskip

A large number of variations of amplitude parameterization, and subsets of fitted energy levels were considered, with particular focus on amplitudes with two $K$-matrix poles and no free constants, and amplitudes that also include a matrix of constants. In each case we used fixed $r^{(0)}, r^{(1)}$ values (which may differ from each other) including $0.1, 0.3, 0.632, 1.0, 3.0$ and $10.0$. The resulting amplitudes feature $t$-matrix poles whose properties are summarized in Figure~\ref{fig:1m_poles}.

We observe that the pole locations show relatively little sensitivity to the choice of amplitude parameterization or $r^{(0)}, r^{(1)}$ value, particularly if only the lowest obtained $\chi^2$ values are retained. The pole residue couplings to the $\etaOctet \etaOctet$ channel are similarly insensitive, and are seen to be of similar magnitude for the two poles. The lighter pole, $\mathsf{a}$, again shows the property that the ratio of pseudoscalar-vector couplings is not well determined, while the sum of their squares is -- based upon the lowest $\chi^2$ values obtained, there might be a slight preference for the ratio being less than unity. For the heavier pole, $\mathsf{b}$, the pseudoscalar-vector couplings are smaller in magnitude, and there appears to be no constraint on their ratio.

\section{Resonance Interpretation}
  \label{sec:Resonances}
 
Having established the resonance content of the octet $J^{PC}=1^{--}, 2^{--}, 3^{--}$ partial waves at the $SU(3)$ flavor point, we proceed to a discussion of their properties, followed by some speculation as to the properties of the corresponding resonances at the physical light quark mass. 
In order to put the resonance properties in physical units, we must set the scale of the lattice. The temporal lattice spacing is obtained by comparing the $\Omega$ baryon mass computed on these lattices to the experimentally measured value, ${a_t^{-1} = m_\Omega^{\mathrm{phys}} / ( a_t m_\Omega^{\mathrm{lat}}) = 4655 \, \mathrm{MeV}}$ with a statistical uncertainty that is irrelevant compared to the statistical errors on the computed quantities.
 
A summary of the $t$-matrix pole properties of resonances\footnote{We give the properties of the pole on the proximal sheet above all relevant thresholds.}, with uncertainties that include the variation over parameterization choice is given below:
 
\begin{equation}
\begin{aligned}
\bm{3^{--}, \, \omegaOctet_3} \nonumber \\    
&\sqrt{s_0} = 1968(6) \pm \tfrac{i}{2} 23(4)\, \mathrm{MeV} 
\nonumber \\[1.2ex]
&\big|c_{\etaOctet \etaOctet} \big| = 223(14)\, \mathrm{MeV} 
\nonumber \\[1.2ex]
&\left( \big|c_{\etaOctet \omegaOctet} \big|^2 + \big|c_{\etaOctet \omegaSinglet} \big|^2 \right)^{1/2}  = 140(28) \, \mathrm{MeV}
 \nonumber \\[1.2ex]
&\Gamma_{\etaOctet \etaOctet} = 18(2)\, \mathrm{MeV}, \;\; \Gamma_{\etaOctet \omegaOctet}+ \Gamma_{\etaOctet \omegaSinglet} = 5(2)\, \mathrm{MeV}
\end{aligned}
\end{equation}

\begin{equation}
\begin{aligned}
\bm{2^{--}, \, \omegaOctet_2} \nonumber \\    
&\sqrt{s_0} = 1966(6) \pm \tfrac{i}{2} 51(7)\, \mathrm{MeV} 
\nonumber \\[1.2ex]
&\left( \big|c_{\etaOctet \omegaOctet \{\threePtwo\}} \big|^2 + \big|c_{\etaOctet \omegaSinglet \{\threePtwo\}} \big|^2 \right)^{1/2}  = 386(51) \, \mathrm{MeV}
 \nonumber \\[1.2ex]
&\left( \big|c_{\etaOctet \omegaOctet \{\threeFtwo\}} \big|^2 + \big|c_{\etaOctet \omegaSinglet \{\threeFtwo\}} \big|^2 \right)^{1/2}  = 233(61) \, \mathrm{MeV}
 \nonumber \\[1.2ex]
\end{aligned}
\end{equation}

\begin{equation}
\begin{aligned}
\bm{1^{--}, \, \omegaOctet_\mathsf{a}} \nonumber \\    
&\sqrt{s_0} = 1786(8) \pm \tfrac{i}{2} 87(11)\, \mathrm{MeV} 
\nonumber \\[1.2ex]
&\big|c_{\etaOctet \etaOctet} \big| = 233(19)\, \mathrm{MeV} 
\nonumber \\[1.2ex]
&\left( \big|c_{\etaOctet \omegaOctet} \big|^2 + \big|c_{\etaOctet \omegaSinglet} \big|^2 \right)^{1/2}  = 587(28) \, \mathrm{MeV}
 \nonumber \\[1.2ex]
&\Gamma_{\etaOctet \etaOctet} = 19(3)\, \mathrm{MeV}, \;\; \Gamma_{\etaOctet \omegaOctet}+ \Gamma_{\etaOctet \omegaSinglet} = 60(6)\, \mathrm{MeV}
\end{aligned}
\end{equation}

\begin{equation}
\begin{aligned}
\bm{1^{--}, \, \omegaOctet_\mathsf{b}} \nonumber \\    
&\sqrt{s_0} = 1957(8) \pm \tfrac{i}{2} 35(8)\, \mathrm{MeV} 
\nonumber \\[1.2ex]
&\big|c_{\etaOctet \etaOctet} \big| = 270(33)\, \mathrm{MeV} 
\nonumber \\[1.2ex]
&\left( \big|c_{\etaOctet \omegaOctet} \big|^2 + \big|c_{\etaOctet \omegaSinglet} \big|^2 \right)^{1/2}  = 177(51) \, \mathrm{MeV}
 \nonumber \\[1.2ex]
&\Gamma_{\etaOctet \etaOctet} = 26(6)\, \mathrm{MeV}, \;\; \Gamma_{\etaOctet \omegaOctet}+ \Gamma_{\etaOctet \omegaSinglet} = 8(5)\, \mathrm{MeV}
\end{aligned}
\end{equation}

In this summary we have made use of the PDG-advocated approach to define partial-widths in terms of pole couplings~\cite{Workman:2022ynf} ($\Gamma_i = \tfrac{\rho_i}{m_R} |c_i|^2$). We observe in each case that the sum of partial widths is in reasonable agreement with the total width defined as twice the imaginary part of the pole position, as one might expect for relatively narrow resonances.

\begin{figure}
\includegraphics[width=\columnwidth]{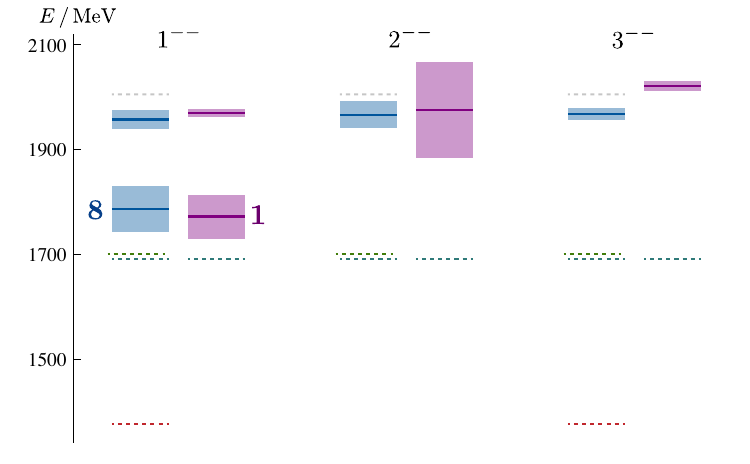}
\caption{Spectrum of octet and singlet resonances at the $SU(3)$ flavor point. A solid line indicate the state's pole mass while the bands show the pole width (uncertainties are not shown). Kinematic thresholds for $\etaOctet \etaOctet$, $\etaOctet \omegaOctet$, $\etaOctet \omegaSinglet$, and $\omegaOctet \omegaOctet$ indicated by the dashed lines.}
\label{fig:spec81}
\end{figure}

\bigskip

Figure~\ref{fig:spec81} compares the spectrum of determined octet resonances with the previously determined spectrum of singlet resonances taken from Ref.~\cite{Johnson:2020ilc}. The only significant mass difference is seen to be between the $3^{--}$ states, where the singlet is somewhat heavier than the octet. The similarity in width of the lighter octet and singlet $1^{--}$ states should be viewed as a coincidence given that the octet state is allowed to decay into $\etaOctet \etaOctet$, a mode forbidden to the singlet, and similarly for the width of the $3^{--}$ states.

We can perform a limited investigation of realization of the OZI-rule in this data by using the common decay mode of both singlet and octet resonances to $\etaOctet \omegaOctet$. Our inability to determine separately the $\etaOctet \omegaOctet$ and $\etaOctet \omegaSinglet$ decay rates (discussed above) restricts what we can achieve here, but we can at least ask if the expectations of ``pure OZI'' as outlined in Ref.~\cite{Johnson:2020ilc} are compatible with the couplings obtained made in this paper. In the ``pure OZI'' limit, the ratio,
\begin{equation}
\left| \frac{g^\mathbf{8}}{g^\mathbf{1}} \right| \equiv \frac{|c(\omega_J^\mathbf{8} \to \etaOctet \omegaOctet)|}{ |c(\omega_J^\mathbf{1} \to \etaOctet \omegaOctet)| } = \frac{\sqrt{5}}{4} \approx 0.56 \, ,
\label{eq:OZI}
\end{equation}
where the notation on the left matches with that in Ref.~\cite{Johnson:2020ilc}.
Using the OZI-rule value of ${c(\omega_J^\mathbf{8} \to \etaOctet \omegaSinglet) / c(\omega_J^\mathbf{8} \to \etaOctet \omegaOctet) = 0.632}$, we obtain the following ratios for the resonances shown in Figure~\ref{fig:spec81}:
\begin{align*}
\frac{|c(\omega_\mathsf{a}^\mathbf{8} \to \etaOctet \omegaOctet)|}{ |c(\omega_\mathsf{a}^\mathbf{1} \to \etaOctet \omegaOctet)| } &= 0.76(6) \, , \\
\frac{|c(\omega_\mathsf{b}^\mathbf{8} \to \etaOctet \omegaOctet)|}{ |c(\omega_\mathsf{b}^\mathbf{1} \to \etaOctet \omegaOctet)| } &= 0.62(30) \, , \\
\frac{|c(\omega_2^\mathbf{8} \to \etaOctet \omegaOctet \{\threePtwo\} )|}{ |c(\omega_2^\mathbf{1} \to \etaOctet \omegaOctet \{\threePtwo\})| } &= 0.40(5) \, , \\
\frac{|c(\omega_2^\mathbf{8} \to \etaOctet \omegaOctet \{\threeFtwo\} )|}{ |c(\omega_2^\mathbf{1} \to \etaOctet \omegaOctet \{\threeFtwo\})| } &= 0.72(21) \, , \\
\frac{|c(\omega_3^\mathbf{8} \to \etaOctet \omegaOctet)|}{ |c(\omega_3^\mathbf{1} \to \etaOctet \omegaOctet)| } &= 0.40(19) \, ,
\end{align*}
and none of these are observed to be in significant tension with the value expected from the OZI-rule.

In summary the only hint of any significant departure from the OZI-rule at the $SU(3)$ flavor point, and hence of a role for disconnected diagrams is the mass of the narrow singlet $3^{--}$ resonance being somewhat larger than the mass of the corresponding octet resonance. Since higher angular momentum states (a $3^{--}$ resonance decays in $F$-waves) have wavefunctions  with more rapid spatial variation, we might expect them to be more sensitive to spatial discretization, so we defer from drawing any strong conclusions about this mass difference until calculations like this one can be performed on lattices with more than one lattice spacing.

We now proceed to consider an attempt to infer some properties of $J^{--}$ resonances at the physical light quark mass based upon the empirical observation from several prior lattice QCD calculations that resonance pole couplings, when scaled by a factor $k^\ell$ reflecting the angular momentum barrier in the decay, appear to be largely independent of the quark mass~\cite{Woss:2020ayi}.

  \subsection{Estimating $J^{--}$ resonance properties at the physical $u,d$ quark mass}
  \label{sec:extrapolate}

Upon breaking the $SU(3)$ flavor symmetry by lowering the (still degenerate) $u,d$ quark masses below the strange quark mass, the degenerate isospin--1 parts of the octet ($\rho^\star$) separate from the degenerate isospin--1/2 parts ($K^\star$), and the previously independent isospin--0 parts of the octet and singlet can admix to form two states, which might be denoted $\omega^\star$, $\phi^\star$.
The use of $\omega$ and $\phi$ as symbols reflects a `lore', based upon assuming the OZI-rule, that the octet and singlet will strongly admix to produce dominantly hidden light-quark $u\bar{u} + d\bar{d}$ ($\omega$), and hidden strange quark $s\bar{s}$ ($\phi$) states.

In Ref.~\cite{Johnson:2020ilc}, partial decay widths into pseudoscalar-vector final states of the $\rho^\star$, $\omega^\star$ and $\phi^\star$ resonances were extrapolated from the \emph{singlet} couplings, assuming a perfect implementation of the OZI-rule.
The calculation reported on in this paper which gives us also the \emph{octet} couplings to pseudoscalar-vector and pseudoscalar-pseudoscalar final states allows us to perform a more complete study.

The physical decay momentum is required to evaluate the scaled couplings, and we currently lack a lattice QCD estimate of the resonance masses at the physical point, so we resort to using experimental masses where available, taken from the PDG averages~\cite{Workman:2022ynf}.

A number of factors limit the predictive power of our approach. Firstly, the already discussed lack of constraint on the ratio of couplings to $\etaOctet \omegaSinglet$ and $\etaOctet \omegaOctet$, which we will assume lies in a region around the OZI-rule value. Secondly, $K^\star$ states are not eigenstates of charge-conjugation (or $G$-parity) and hence a $J^P$ state can in general be an admixture of $J^{P-}$ and $J^{P+}$ basis states, the latter of which we have not computed in this study\footnote{Since $1^{-+}$ and $3^{-+}$ are exotic in the sense of being inaccessible to a $q\bar{q}$ pair, and because lightest hybrid mesons with these quantum numbers lie at much heavier mass~\cite{Dudek:2009qf, Dudek:2010wm, Dudek:2011bn, Dudek:2013yja}, we expect them to be negligible components in the $K^*$, $K^*_3$ resonances.}. Finally, the degree of admixture of octet and singlet basis states to form the $\omega^\star$ and $\phi^\star$ eigenstates is completely unknown to us using only a calculation at the $SU(3)$ flavor point, where octet and singlet are independent representations. In this last case, we will indicate the sensitivity of our results to the value of the octet-singlet mixing angle.

\bigskip

As in Ref.~\cite{Johnson:2020ilc}, we make use of decomposition of elements of the $SU(3)$ representations into states labelled by isospin and strangeness~\cite{deSwart:1963pdg, Woss:2020ayi}. For example, for pseudoscalar-pseudoscalar decays of the octet,
{\scriptsize
\begin{align*}
\big| \mathbf{8}_2&; S=0, I=1, I_z=+1 \big\rangle \\
&= \sqrt{\tfrac{1}{6}}\left( K^+ \overline{K}^0 - \overline{K}^0 K^+ \right) + \sqrt{\tfrac{1}{3}} \left( \pi^+ \pi^0 - \pi^0 \pi^+ \right) \\[1.4ex]
\big| \mathbf{8}_2&; S=1, I=\tfrac{1}{2}, I_z=+\tfrac{1}{2} \big\rangle \\
&= \sqrt{\tfrac{1}{12}}\left( K^+ \pi^0 - \pi^0 K^+ \right) - \sqrt{\tfrac{1}{6}} \left( K^0 \pi^+ - \pi^+ K^0 \right) + \tfrac{1}{2}\left( K^+ \eta_8 - \eta_8 K^+ \right) \\[1.4ex]
\big| \mathbf{8}_2&; S=0, I=0, I_z=0 \big\rangle \\
&= \tfrac{1}{2}\left( K^+ K^- - K^- K^+ - K^0 \overline{K}^0 +\overline{K}^0 K^0  \right) \, ,
\end{align*}
}
\!\!\!which will be used to evaluate decays of the $\rho^\star$, $K^\star$ and $\omega^\star/\phi^\star$ resonances respectively.

In cases of resonance decays to final states containing the $\eta$, we will make use of the empirical observation that the $\eta$ is very close to being an $SU(3)$ flavor octet state\footnote{See also Ref.~\cite{Dudek:2013yja} for determinations of the mixing on lattices related to those used in this paper.}, and set $\eta = \eta_8$.
In decays to final states containing the $\omega(782)$ or the $\phi(1020)$, we will treat these as ideally flavor mixed so that,
\begin{align*}
\omega = \tfrac{1}{\sqrt{2}} \left( u\bar{u} + d\bar{d} \right) &= \sqrt{\tfrac{1}{3}} \, \omega_8 + \sqrt{\tfrac{2}{3}}\,  \omega_1 \\
\phi = s\bar{s} &= -\sqrt{\tfrac{2}{3}}\,  \omega_8 + \sqrt{\tfrac{1}{3}}\,  \omega_1 \, ,
\end{align*}
which naturally introduces sensitivity to the ratio of couplings to $\etaOctet \omegaSinglet$ and $\etaOctet \omegaOctet$.

We will compute partial widths for pseudoscalar-pseudoscalar and pseudoscalar-vector final states, and then sum these to estimate the total hadronic width, but since we cannot predict partial widths into channels which are kinematically closed at the $SU(3)$ flavor point (as we have no constraint on their couplings), generally we will predict lower limits on the total width of each resonance.
In the $1^{--}$ sector, we will assume that the two states present at the $SU(3)$ point evolve into two states (in each flavor channel) at the physical point without mixing with each other.
We will not propagate the statistical errors on the $SU(3)$ flavor point couplings to avoid the appearance of any kind of quantified uncertainty on these (potentially crude) model-dependent predictions.
We will present only a limited phenomenological comparison with experimental data, reserving a more complete discussion for a future publication.

\subsubsection{$\rho^\star_J$ resonances}

The isovector resonances depend only upon the octet couplings presented in the current paper. We scale them as described above, e.g.
\begin{equation*}
g(\rho^\star_J \to \pi \pi) = \sqrt{\frac{2}{3}} \, \big| c(\omega^\mathbf{8}_J \to \etaOctet \etaOctet) \big| \,  \left|\frac{k_\mathrm{phys}}{k_\mathrm{lat}}\right|^\ell \, ,
\end{equation*}
where for pseudoscalar-pseudoscalar decays, $\ell = J$, and where we are considering the coupling to include the sum over possible final state charge configurations. Using this coupling, the corresponding partial-width can be obtained,
\begin{equation*}
\Gamma(\rho^\star_J \to \pi \pi) = \frac{\rho_\mathrm{phys}}{m_\mathrm{phys}} g(\rho^\star_J \to \pi \pi)^2 \, .
\end{equation*}

For resonance decays to pseudoscalar-vector final states, our lack of constraint on the $\etaOctet \omegaSinglet$, $\etaOctet \omegaOctet$ coupling ratio is expressed in relationships like,
\begin{align*}
g(\rho^\star_J \to \pi \omega) &= \sqrt{\frac{1}{15}} \frac{ 1 + \sqrt{10} \, r}{\sqrt{1+ r^2}}\,  \tilde{c}\, , \\
g(\rho^\star_J \to \pi \phi) &= -\sqrt{\frac{2}{15}} \frac{ 1 - \sqrt{5/2} \, r}{\sqrt{1+ r^2}}\,  \tilde{c} \, ,
\end{align*}
where $\tilde{c}$ is the well-constrained combination $\left( |c_{\etaOctet \omegaOctet}|^2 + |c_{\etaOctet \omegaSinglet}|^2 \right)^{1/2}$, while $r$ is the poorly constrained ratio, $|c_{\etaOctet \omegaSinglet}| / |c_{\etaOctet \omegaOctet}|$. Notice that in the limit $r \to \sqrt{2/5} \approx 0.632 $, the decay $\rho^\star_J \to \pi \phi$ has zero coupling, in line with the OZI-rule forbidding production of the completely disconnected $\phi = s\bar{s}$ in this decay.

\bigskip
\noindent$\bm{\rho_3(1690)}$ -- With $r=0.632$, the $F$-wave decays of this state extrapolate to

\smallskip
\begin{tabular}{cccccc|l}
$\pi\pi$ & $K\overline{K}$ & $\pi \omega$ & $K\overline{K}^*$ & $\eta \rho$ & $\pi \phi$ & total \\
53       & 7               & 20           & 2                 & 1           & 0          & 83  \,\, MeV \, .
\end{tabular}
\smallskip

\noindent The $\pi \omega$ decay is the only relevant mode to have significant sensitivity to the value of $r$, becoming close to zero for small value of $r$, and reaching a maximum near \mbox{35 MeV} for $r \sim 3$. As such the predicted total width is relatively insensitive to $r$, not exceeding 100 MeV, which is somewhat lower than the PDG average width, although the PDG reports considerable variation in width between analysis of different final states. The significant experimental branch into $\pi\pi\pi\pi$ (excluding $\pi \omega$) would not be captured in our extrapolation, so an under estimate of width may be appropriate.

\medskip
\noindent$\bm{\rho_2(1690)}$ -- Given the absence of an experimental candidate for this state, we set the mass equal to the PDG mass of the $\rho_3$ considered above. The extrapolated partial-widths (summing $P$-wave and $F$-wave contributions) for $r=0.632$ are

\smallskip
\begin{tabular}{cccc|l}
$\pi \omega$ & $K\overline{K}^*$ & $\eta \rho$ & $\pi \phi$ & total \\
108          & 24                & 13          & 0          & 145  \,\, MeV \, .
\end{tabular}
\smallskip

\noindent Again, the $\pi \omega$ mode shows sensitivity to the chosen value of $r$, and a non-negligible branch into $\pi \phi$ is generated away from the OZI-rule point, but this partial-width does not get above 10 MeV for $r \lesssim 3$. Because only pseudoscalar-vector decays are present here, the total width is more sensitive to the value of $r$, being as low as 80 MeV for small $r$ or as large as 200 MeV for $ r \sim 3$. The possibility of a kinematically open $S$-wave decay into $\pi a_2$ (which will dominantly populate a $\pi\pi\pi\pi$ final state), could lead to a much larger total width for this state, and might explain why it has not yet been seen experimentally.

\medskip
\noindent$\bm{\rho(1450)}$ -- Using the PDG average mass of 1465 MeV, for $r=0.632$ we predict

\smallskip
\begin{tabular}{cccccc|l}
$\pi\pi$ & $K\overline{K}$ & $\pi \omega$ & $K\overline{K}^*$ & $\eta \rho$ & $\pi \phi$ & total \\
39       & 8               & 230          & 20                & 18           & 0         & 314  \,\, MeV \, .
\end{tabular}
\smallskip

\noindent The rather large total width of this state is in line with the PDG summary which suggests a total width near \mbox{400 MeV}. The width is dominated by the $\pi \omega$ mode, which is sensitive to the choice of $r$, with significantly smaller total width at smaller values of $r$, and a total width near 450 MeV for larger values of $r$.

\medskip
\noindent$\bm{\rho(1700)}$ -- Using the PDG average mass of 1720 MeV, for $r=0.632$ we predict

\smallskip
\begin{tabular}{cccccc|l}
$\pi\pi$ & $K\overline{K}$ & $\pi \omega$ & $K\overline{K}^*$ & $\eta \rho$ & $\pi \phi$ & total \\
41       & 12              & 12           & 5                 & 2           & 0         & 72  \,\, MeV \, .
\end{tabular}
\smallskip

\noindent This prediction of a relatively narrow resonance is in poor agreement with experiment, where a much broader state is required, although a relatively small rate into $\pi \omega$ was suggested in the analysis of experimental data presented in Ref.~\cite{Clegg:1993mt}. The large mass of this state might suggest that there are additional decay modes kinematically accessible at the physical point not accounted for in our calculation, and these could be significant contributions to the total width.

\subsubsection{$K^\star_J$ resonances}

Our knowledge of the excited kaon resonances comes mostly from the LASS experiment, which published results on $K\pi, K\pi\pi, K\eta \ldots$ final states using a low-energy kaon beam in the early 1980s. One particularly relevant result of theirs is the claim that data in the $K\omega$ final state is best described in terms of \emph{two} $J^P = 2^-$ states~\cite{Aston:1993qc} -- within a $q\bar{q}$ quark model these would correspond to admixtures of the $\threeDtwo (2^{--}), \oneDtwo (2^{-+})$ basis states. 

Our kaon resonance predictions depend only on the octet couplings computed in this paper.

\medskip
\noindent$\bm{K^*_3(1780)}$ -- With $r=0.632$, the $F$-wave decays of this state extrapolate to

\smallskip
\begin{tabular}{ccccccc|l}
$K\pi$ & $K\eta$ & $K \rho$ & $K^* \pi$ & $K^* \eta$ & $K \omega$ & $K\phi$ &  total \\
30     & 13      & 9        & 14        & 0          & 3          & 1       & 70  \,\, MeV \, .
\end{tabular}
\smallskip

\noindent This small total width is well below the PDG width of \mbox{161 MeV}, and variation of the value of $r$ cannot resolve the discrepancy. There remains the possibility of relatively large $P$-wave vector-vector decays whose strength we have not estimated.

\medskip
\noindent$\bm{K_2(1779)}$ -- Since we have no constraint on possible mixing between a $2^{--}$ basis state and a $2^{-+}$ basis state, we will proceed assuming that one $2^-$ kaon resonance is dominantly the $2^{--}$ state we've computed. With ${r=0.632}$ we have

\smallskip
\begin{tabular}{ccccc|l}
$K \rho$ & $K^* \pi$ & $K^* \eta$ & $K \omega$ & $K\phi$ &  total \\
55       & 74        & 3          & 17         & 9       & 158  \,\, MeV \, .
\end{tabular}
\smallskip

\noindent Variation of $r$ changes this total width only by about 20\% up or down. The lighter of the two $2^-$ states in the LASS analysis~\cite{Aston:1993qc} has a total width of roughly this size.

\medskip
\noindent$\bm{K^*(1410)}$ -- With $r=0.632$ this suprisingly low-mass state (it is lighter than the $\rho(1450)$ despite needing to contain a strange quark) has partial widths

\smallskip
\begin{tabular}{ccccccc|l}
$K\pi$ & $K\eta$ & $K \rho$ & $K^* \pi$ & $K^* \eta$ & $K \omega$ & $K\phi$ &  total \\
19     & 9       & 41       & 95        & 0          & 12         & 0       & 176  \,\, MeV \, .
\end{tabular}
\smallskip

\noindent This predicted total width is only slightly below the PDG average for this state, and the LASS results have this state seen in $K^* \pi$ but not in $K\rho$, somewhat supporting the hierarchy of partial widths coming from broken $SU(3)$ flavor symmetry. The total width of this state grows somewhat for smaller values of $r$.

\medskip
\noindent$\bm{K^*(1680)}$ -- Using the PDG average mass of 1718 MeV, and $r=0.632$ we have

\smallskip
\begin{tabular}{ccccccc|l}
$K\pi$ & $K\eta$ & $K \rho$ & $K^* \pi$ & $K^* \eta$ & $K \omega$ & $K\phi$ &  total \\
24     & 16      & 5        & 7         & 0          & 2         & 1       & 56  \,\, MeV \, .
\end{tabular}
\smallskip

\noindent As was the case in the isovector channel, the small value of the $SU(3)$--limit coupling to pseudoscalar-vector causes this resonance to be rather narrow, much narrower than the rather broad experimental state. The total width is largely independent of the value of $r$. The possibility of a large $S$-wave $\pi K_1(1270)$ branch can be considered as an explanation of the discrepancy.

\subsubsection{Isoscalar $\omega^\star_J, \phi^\star_J$ resonances}

\begin{figure*}
\includegraphics[width=0.8\textwidth]{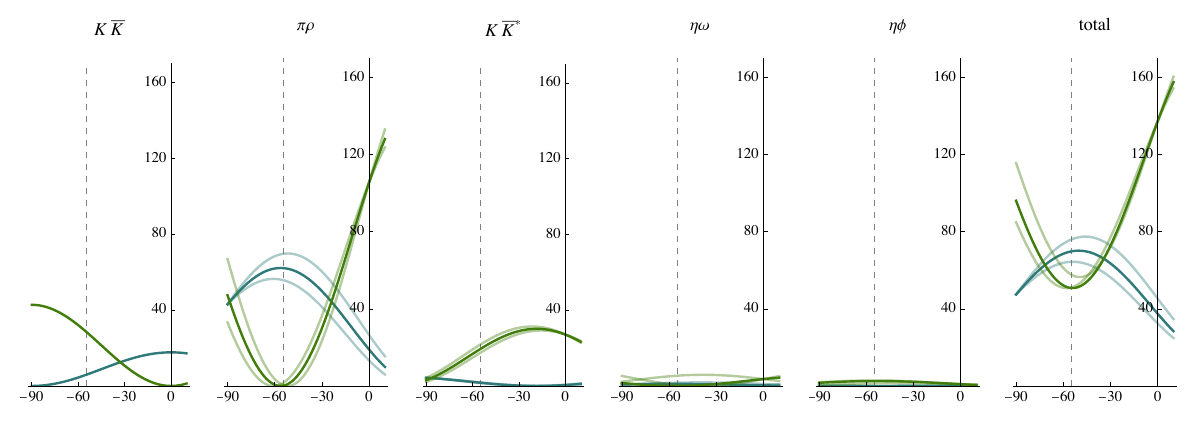}
\includegraphics[width=0.8\textwidth]{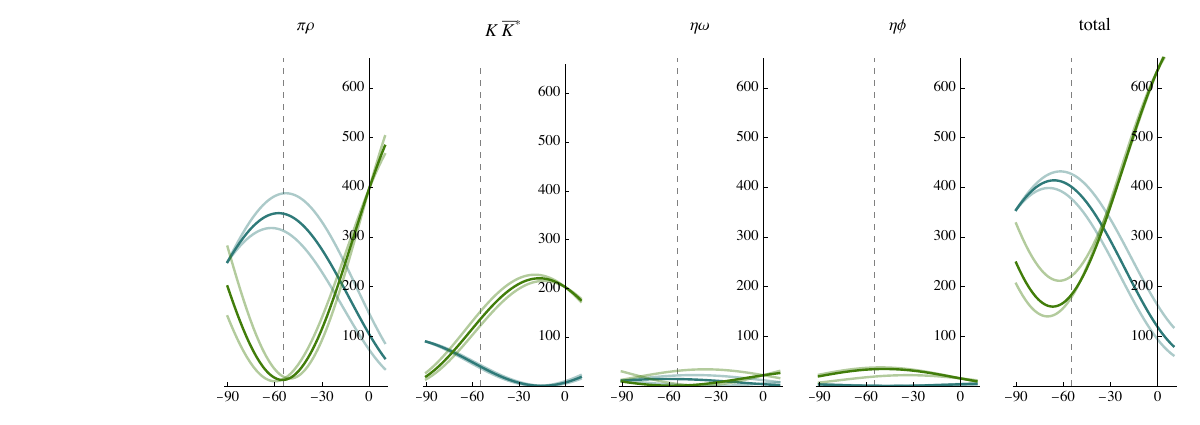}
\includegraphics[width=0.8\textwidth]{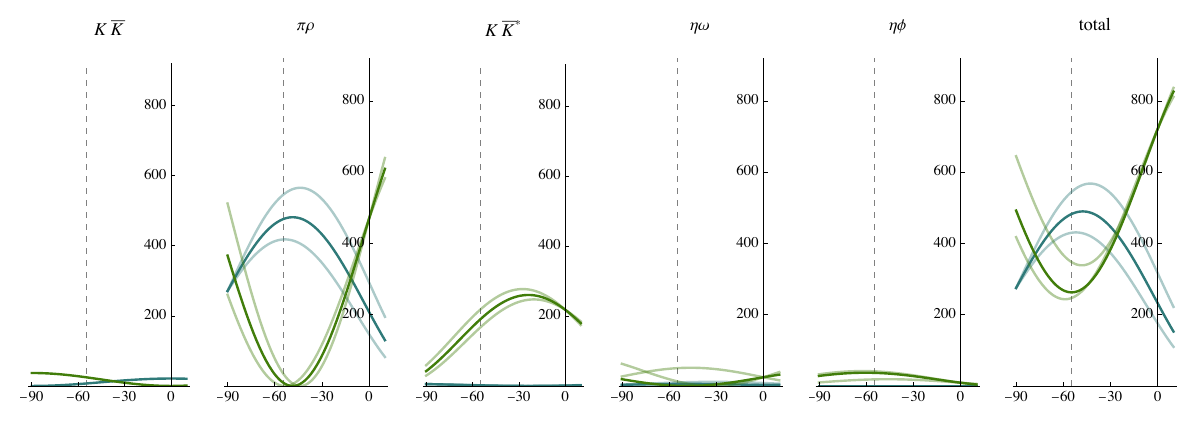}
\includegraphics[width=0.8\textwidth]{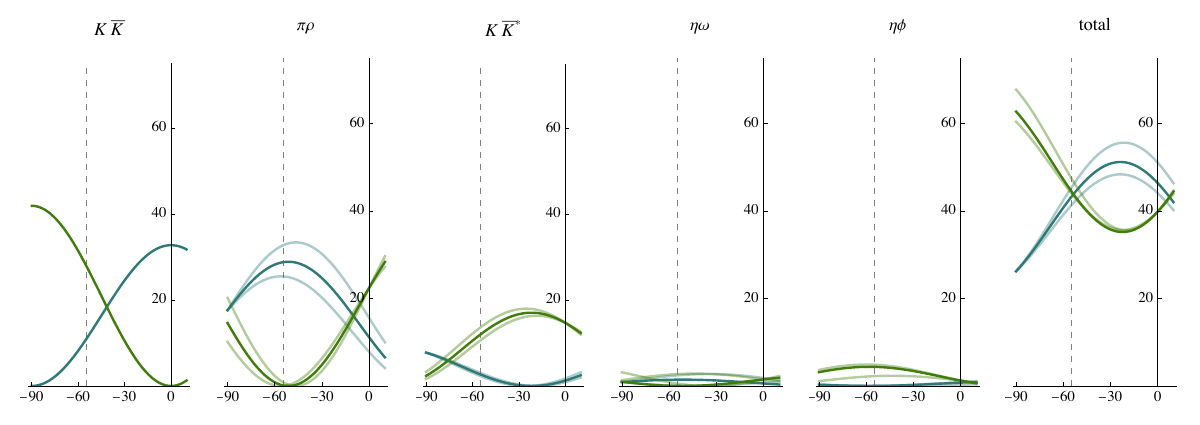}
\caption{Partial widths (in MeV) of $\omega_J$(blue) and $\phi_J$(green) resonances with masses given in the text, as a function of the octet-singlet mixing angle $\theta$. In each case the darker curve corresponds to $r=0.632$ while the two lighter curves correspond to $r=0$ and $r=1$. Top to bottom panels present $3^{--}$, $2^{--}$, $1^{--}_\mathsf{a}$,  $1^{--}_\mathsf{b}$. }
\label{fig:omega_phi}
\end{figure*}

Given our current lack of any constraint on the mixing of $SU(3)$ flavor octet and singlet components to form the physical eigenstates away from the $SU(3)$ flavor point, we will present results as a function of the mixing angle, $\theta$, which appears in,
\begin{align*}
\omega^\star &= \cos \theta \, \omega_8 - \sin \theta \, \omega_1 \, ,\\
\phi^\star   &= \sin \theta \, \omega_8 + \cos \theta \, \omega_1 \, ,
\end{align*}
where the ideal flavor mixing case $\omega^\star \sim u\bar{u} + d\bar{d}$, ${\phi^\star \sim s\bar{s}}$ corresponds to $\theta = -54.74^\circ$. The sensitivity of our partial-width predictions to this mixing angle, and to the value of $r$, is presented in Figure~\ref{fig:omega_phi}.

\pagebreak
\noindent$\bm{\omega_3(1670), \phi_3(1850)}$ -- Using the PDG masses for these states, with $r=0.632$ and the ideal flavor mixing angle (i.e. assuming perfect OZI-rule) we obtain partial-widths

\smallskip
\begin{tabular}{r|ccccc|l}
           & $K\overline{K}$ & $\pi\rho$ & $K\overline{K}^*$ & $\eta \omega$ & $\eta\phi$ &  total \\
$\omega_3$ & 6               & 61        & 2                 & 1             & 0          & 70  \,\, MeV \\
$\phi_3$   & 29              & 0         & 20                & 0             & 3          & 51 \,\, MeV \, .
\end{tabular}
\smallskip

\noindent The sensitivity to $r$ and the mixing angle, $\theta$, is presented in the top row of Figure~\ref{fig:omega_phi}, where the largest sensitivity is seen to be in the partial widths of these two resonances to the $\pi \rho$ final-state.

\medskip
\noindent$\bm{\omega_2(1667), \phi_2(1854)}$ -- Given the lack of any experimental candidate states in the $2^{--}$ isoscalar sector, we choose to reuse the $3^{--}$ masses from above. With $r=0.632$ and the ideal flavor mixing angle we obtain partial-widths (summed over $F$-wave and $P$-wave decays),

\smallskip
\begin{tabular}{r|cccc|l}
           & $\pi\rho$ & $K\overline{K}^*$ & $\eta \omega$ & $\eta\phi$ &  total \\
$\omega_2$ & 347       & 40                & 14            & 0          & 401  \,\, MeV \\
$\phi_2$   & 12        & 137               & 0             & 34         & 183  \,\, MeV \, .
\end{tabular}
\smallskip

\noindent Clearly this extrapolation predicts a very large width for the $\omega_2$ state, even before any possible $S$-wave decays are considered. The sensitivity to $r$ and the mixing angle, $\theta$ is presented in the second row of Figure~\ref{fig:omega_phi}, where it is clear that for any choice of mixing angle, at least one of $\omega_2, \phi_2$ must be broad.

\medskip
\noindent$\bm{\omega(1410), \phi(1680)}$ -- using PDG masses and with ${r=0.632}$ and the ideal flavor mixing angle we obtain partial-widths

\smallskip
\begin{tabular}{r|ccccc|l}
           & $K\overline{K}$ & $\pi\rho$ & $K\overline{K}^*$ & $\eta \omega$ & $\eta\phi$ &  total \\
$\omega$   & 7               & 468       & 2                 & 6             & 0          & 482  \,\, MeV \\
$\phi$     & 25              & 10        & 191               & 0             & 37         & 263  \,\, MeV \, .
\end{tabular}
\smallskip

\noindent The sensitivity to $r$ and the mixing angle, $\theta$ is presented in the third row of Figure~\ref{fig:omega_phi}, where it is clear that for any value of the mixing angle, both light vector meson resonances should be broad.
The $\omega(1410)$ resonance is very poorly determined experimentally, with the PDG average for its total width being computed from completely incompatible extractions varying from 100 MeV to 880 MeV. 
The PDG average of $e^+e^-$ results has the $\phi(1680)$ with a width of only $150(50)$ MeV, well below our prediction, but examination of the data entering this average includes results like those in Ref.~\cite{BaBar:2007ceh}, where $e^+e^- \to K\overline{K}^*, \eta \phi$ isoscalar cross-sections feature a single broad bump described as a resonance with a width over 300 MeV.

\medskip
\noindent$\bm{\omega(1670), \phi(1850)}$ -- The mass used for the $\omega$ state corresponds to the PDG average for a state claimed in multiple experiments, while the $\phi$ mass is based on the $\phi_3$ mass, as there is currently no second $\phi$ resonance candidate. Using $r=0.632$ and the ideal flavor mixing angle we obtain partial-widths

\smallskip
\begin{tabular}{r|ccccc|l}
           & $K\overline{K}$ & $\pi\rho$ & $K\overline{K}^*$ & $\eta \omega$ & $\eta\phi$ &  total \\
$\omega$   & 11              & 28        & 3                 & 1             & 0          & 43  \,\, MeV \\
$\phi$     & 28              & 0         & 12                & 0             & 4          & 44  \,\, MeV \, .
\end{tabular}
\smallskip

\noindent The sensitivity to $r$ and the mixing angle, $\theta$ is presented in the bottom row of Figure~\ref{fig:omega_phi}, where we see that the decays accounted for in this analysis cannot generate large widths for these states. 
As was the case for the $\omega(1410)$, the $\omega(1670)$ width varies considerably between experiments, but it is certainly significantly larger than any of the values presented in the bottom right panel of Figure~\ref{fig:omega_phi}.

\section{Summary}
  \label{sec:Summary}
 
We have presented a first calculation within lattice QCD of the $1^{--}$, $2^{--}$ and $3^{--}$ coupled-channel scattering amplitudes in the octet representation of $SU(3)$ flavor, finding relatively narrow resonances in each case. We demonstrated that the near degeneracy of the stable $\omegaOctet$ and $\omegaSinglet$ mesons severely limits our ability to determine independently the $\etaOctet \omegaOctet$ and $\etaOctet \omegaSinglet$ couplings of these resonances, but that the sum of these two modes is well determined, as is the coupling to $\etaOctet \etaOctet$ (when allowed by symmetries). The obtained results are broadly compatible with the accepted ``OZI-rule'' phenomenology in which disconnected quark-line diagrams are suppressed.

The results of this paper, taken together with earlier results on the $SU(3)$ flavor \emph{singlet} channel, were used to perform a speculative extrapolation to the physical light quark mass, in order to make predictions for the properties of $\rho_J^\star, K_J^\star, \omega_J^\star$ and $\phi_J^\star$ resonances. In order to make more reliable QCD predictions without undue model-dependence it will be necessary to perform lattice calculations with lower values of the light quark mass, explicitly breaking the $SU(3)$ flavor symmetry. This will lead to a number of complications such as the growth in the number of scattering channels (e.g. $\etaOctet \etaOctet \to \pi\pi, K\overline{K}, \eta\eta$) and the presence of three-meson scattering channels in the energy regions of interest~\cite{Hansen:2020otl}.

\begin{acknowledgments}
We thank our colleagues within the Hadron Spectrum Collaboration, in particular D.J.~Wilson for his assistance with the analysis code and C.E.~Thomas for comments on an earlier version of the manuscript. JJD and CTJ acknowledge support from the U.S. Department of Energy contract DE-SC0018416 at William \& Mary, and contract DE-AC05-06OR23177, under which Jefferson Science Associates, LLC, manages and operates Jefferson Lab. 
This work contributes to the goals of the U.S. Department of Energy \emph{ExoHad} Topical Collaboration, Contract No. DE-SC0023598.

The software codes
{\tt Chroma}~\cite{Edwards:2004sx} and {\tt QUDA}~\cite{Clark:2009wm,Babich:2010mu,Clark:2016rdz} were used. 
The authors acknowledge support from the U.S. Department of Energy, Office of Science, Office of Advanced Scientific Computing Research and Office of Nuclear Physics, Scientific Discovery through Advanced Computing (SciDAC) program. 
Also acknowledged is support from the Exascale Computing Project (17-SC-20-SC), a collaborative effort of the U.S. Department of Energy Office of Science and the National Nuclear Security Administration.
This work was performed using the Cambridge Service for Data Driven Discovery (CSD3) operated by the University of Cambridge Research Computing Service (www.hpc.cam.ac.uk), provided by Dell EMC and Intel using Tier-2 funding from the Engineering and Physical Sciences Research Council (capital grant EP/P020259/1), and DiRAC funding from STFC (www.dirac.ac.uk). The DiRAC component of CSD3 was funded by BEIS capital funding via STFC capital grants ST/P002307/1 and ST/R002452/1 and STFC operations grant ST/R00689X/1. DiRAC is part of the National e-Infrastructure.
This work was also performed on clusters at Jefferson Lab under the USQCD Collaboration and the LQCD ARRA Project.
This research was supported in part under an ALCC award, and used resources of the Oak Ridge Leadership Computing Facility at the Oak Ridge National Laboratory, which is supported by the Office of Science of the U.S. Department of Energy under Contract No. DE-AC05-00OR22725.
This research used resources of the National Energy Research Scientific Computing Center (NERSC), a DOE Office of Science User Facility supported by the Office of Science of the U.S. Department of Energy under Contract No. DE-AC02-05CH11231.
The authors acknowledge the Texas Advanced Computing Center (TACC) at The University of Texas at Austin for providing HPC resources.
Gauge configurations were generated using resources awarded from the U.S. Department of Energy INCITE program at the Oak Ridge Leadership Computing Facility, the NERSC, the NSF Teragrid at the TACC and the Pittsburgh Supercomputer Center, as well as at the Cambridge Service for Data Driven Discovery (CSD3) and Jefferson Lab.
This work was performed in part using computing facilities at William \& Mary which were provided by contributions from the National Science Foundation (MRI grant PHY-1626177), and the Commonwealth of Virginia Equipment Trust Fund.
\end{acknowledgments}



\bibliographystyle{apsrev4-1}
\bibliography{bib}

\begin{thebibliography}{42}%
\makeatletter
\providecommand \@ifxundefined [1]{%
 \@ifx{#1\undefined}
}%
\providecommand \@ifnum [1]{%
 \ifnum #1\expandafter \@firstoftwo
 \else \expandafter \@secondoftwo
 \fi
}%
\providecommand \@ifx [1]{%
 \ifx #1\expandafter \@firstoftwo
 \else \expandafter \@secondoftwo
 \fi
}%
\providecommand \natexlab [1]{#1}%
\providecommand \enquote  [1]{``#1''}%
\providecommand \bibnamefont  [1]{#1}%
\providecommand \bibfnamefont [1]{#1}%
\providecommand \citenamefont [1]{#1}%
\providecommand \href@noop [0]{\@secondoftwo}%
\providecommand \href [0]{\begingroup \@sanitize@url \@href}%
\providecommand \@href[1]{\@@startlink{#1}\@@href}%
\providecommand \@@href[1]{\endgroup#1\@@endlink}%
\providecommand \@sanitize@url [0]{\catcode `\\12\catcode `\$12\catcode
  `\&12\catcode `\#12\catcode `\^12\catcode `\_12\catcode `\%12\relax}%
\providecommand \@@startlink[1]{}%
\providecommand \@@endlink[0]{}%
\providecommand \url  [0]{\begingroup\@sanitize@url \@url }%
\providecommand \@url [1]{\endgroup\@href {#1}{\urlprefix }}%
\providecommand \urlprefix  [0]{URL }%
\providecommand \Eprint [0]{\href }%
\providecommand \doibase [0]{http://dx.doi.org/}%
\providecommand \selectlanguage [0]{\@gobble}%
\providecommand \bibinfo  [0]{\@secondoftwo}%
\providecommand \bibfield  [0]{\@secondoftwo}%
\providecommand \translation [1]{[#1]}%
\providecommand \BibitemOpen [0]{}%
\providecommand \bibitemStop [0]{}%
\providecommand \bibitemNoStop [0]{.\EOS\space}%
\providecommand \EOS [0]{\spacefactor3000\relax}%
\providecommand \BibitemShut  [1]{\csname bibitem#1\endcsname}%
\let\auto@bib@innerbib\@empty
\bibitem [{\citenamefont {Shepherd}\ \emph {et~al.}(2016)\citenamefont
  {Shepherd}, \citenamefont {Dudek},\ and\ \citenamefont
  {Mitchell}}]{Shepherd:2016dni}%
  \BibitemOpen
  \bibfield  {author} {\bibinfo {author} {\bibfnamefont {M.~R.}\ \bibnamefont
  {Shepherd}}, \bibinfo {author} {\bibfnamefont {J.~J.}\ \bibnamefont {Dudek}},
  \ and\ \bibinfo {author} {\bibfnamefont {R.~E.}\ \bibnamefont {Mitchell}},\
  }\href {\doibase 10.1038/nature18011} {\bibfield  {journal} {\bibinfo
  {journal} {Nature}\ }\textbf {\bibinfo {volume} {534}},\ \bibinfo {pages}
  {487} (\bibinfo {year} {2016})},\ \Eprint {http://arxiv.org/abs/1802.08131}
  {arXiv:1802.08131 [hep-ph]} \BibitemShut {NoStop}%
\bibitem [{\citenamefont {Johnson}\ and\ \citenamefont
  {Dudek}(2021)}]{Johnson:2020ilc}%
  \BibitemOpen
  \bibfield  {author} {\bibinfo {author} {\bibfnamefont {C.~T.}\ \bibnamefont
  {Johnson}}\ and\ \bibinfo {author} {\bibfnamefont {J.~J.}\ \bibnamefont
  {Dudek}} (\bibinfo {collaboration} {Hadron Spectrum}),\ }\href {\doibase
  10.1103/PhysRevD.103.074502} {\bibfield  {journal} {\bibinfo  {journal}
  {Phys. Rev. D}\ }\textbf {\bibinfo {volume} {103}},\ \bibinfo {pages}
  {074502} (\bibinfo {year} {2021})},\ \Eprint
  {http://arxiv.org/abs/2012.00518} {arXiv:2012.00518 [hep-lat]} \BibitemShut
  {NoStop}%
\bibitem [{\citenamefont {Luscher}(1986{\natexlab{a}})}]{Luscher:1985dn}%
  \BibitemOpen
  \bibfield  {author} {\bibinfo {author} {\bibfnamefont {M.}~\bibnamefont
  {Luscher}},\ }\href {\doibase 10.1007/BF01211589} {\bibfield  {journal}
  {\bibinfo  {journal} {Commun. Math. Phys.}\ }\textbf {\bibinfo {volume}
  {104}},\ \bibinfo {pages} {177} (\bibinfo {year}
  {1986}{\natexlab{a}})}\BibitemShut {NoStop}%
\bibitem [{\citenamefont {Luscher}(1986{\natexlab{b}})}]{Luscher:1986pf}%
  \BibitemOpen
  \bibfield  {author} {\bibinfo {author} {\bibfnamefont {M.}~\bibnamefont
  {Luscher}},\ }\href {\doibase 10.1007/BF01211097} {\bibfield  {journal}
  {\bibinfo  {journal} {Commun. Math. Phys.}\ }\textbf {\bibinfo {volume}
  {105}},\ \bibinfo {pages} {153} (\bibinfo {year}
  {1986}{\natexlab{b}})}\BibitemShut {NoStop}%
\bibitem [{\citenamefont {Briceno}\ \emph
  {et~al.}(2018{\natexlab{a}})\citenamefont {Briceno}, \citenamefont {Dudek},\
  and\ \citenamefont {Young}}]{Briceno:2017max}%
  \BibitemOpen
  \bibfield  {author} {\bibinfo {author} {\bibfnamefont {R.~A.}\ \bibnamefont
  {Briceno}}, \bibinfo {author} {\bibfnamefont {J.~J.}\ \bibnamefont {Dudek}},
  \ and\ \bibinfo {author} {\bibfnamefont {R.~D.}\ \bibnamefont {Young}},\
  }\href {\doibase 10.1103/RevModPhys.90.025001} {\bibfield  {journal}
  {\bibinfo  {journal} {Rev. Mod. Phys.}\ }\textbf {\bibinfo {volume} {90}},\
  \bibinfo {pages} {025001} (\bibinfo {year} {2018}{\natexlab{a}})},\ \Eprint
  {http://arxiv.org/abs/1706.06223} {arXiv:1706.06223 [hep-lat]} \BibitemShut
  {NoStop}%
\bibitem [{\citenamefont {Edwards}\ \emph {et~al.}(2008)\citenamefont
  {Edwards}, \citenamefont {Joo},\ and\ \citenamefont {Lin}}]{Edwards:2008ja}%
  \BibitemOpen
  \bibfield  {author} {\bibinfo {author} {\bibfnamefont {R.~G.}\ \bibnamefont
  {Edwards}}, \bibinfo {author} {\bibfnamefont {B.}~\bibnamefont {Joo}}, \ and\
  \bibinfo {author} {\bibfnamefont {H.-W.}\ \bibnamefont {Lin}},\ }\href
  {\doibase 10.1103/PhysRevD.78.054501} {\bibfield  {journal} {\bibinfo
  {journal} {Phys. Rev.}\ }\textbf {\bibinfo {volume} {D78}},\ \bibinfo {pages}
  {054501} (\bibinfo {year} {2008})},\ \Eprint {http://arxiv.org/abs/0803.3960}
  {arXiv:0803.3960 [hep-lat]} \BibitemShut {NoStop}%
\bibitem [{\citenamefont {Lin}\ \emph {et~al.}(2009)\citenamefont {Lin} \emph
  {et~al.}}]{Lin:2008pr}%
  \BibitemOpen
  \bibfield  {author} {\bibinfo {author} {\bibfnamefont {H.-W.}\ \bibnamefont
  {Lin}} \emph {et~al.} (\bibinfo {collaboration} {Hadron Spectrum}),\ }\href
  {\doibase 10.1103/PhysRevD.79.034502} {\bibfield  {journal} {\bibinfo
  {journal} {Phys. Rev.}\ }\textbf {\bibinfo {volume} {D79}},\ \bibinfo {pages}
  {034502} (\bibinfo {year} {2009})},\ \Eprint {http://arxiv.org/abs/0810.3588}
  {arXiv:0810.3588 [hep-lat]} \BibitemShut {NoStop}%
\bibitem [{\citenamefont {Peardon}\ \emph {et~al.}(2009)\citenamefont
  {Peardon}, \citenamefont {Bulava}, \citenamefont {Foley}, \citenamefont
  {Morningstar}, \citenamefont {Dudek}, \citenamefont {Edwards}, \citenamefont
  {Joo}, \citenamefont {Lin}, \citenamefont {Richards},\ and\ \citenamefont
  {Juge}}]{Peardon:2009gh}%
  \BibitemOpen
  \bibfield  {author} {\bibinfo {author} {\bibfnamefont {M.}~\bibnamefont
  {Peardon}}, \bibinfo {author} {\bibfnamefont {J.}~\bibnamefont {Bulava}},
  \bibinfo {author} {\bibfnamefont {J.}~\bibnamefont {Foley}}, \bibinfo
  {author} {\bibfnamefont {C.}~\bibnamefont {Morningstar}}, \bibinfo {author}
  {\bibfnamefont {J.}~\bibnamefont {Dudek}}, \bibinfo {author} {\bibfnamefont
  {R.~G.}\ \bibnamefont {Edwards}}, \bibinfo {author} {\bibfnamefont
  {B.}~\bibnamefont {Joo}}, \bibinfo {author} {\bibfnamefont {H.-W.}\
  \bibnamefont {Lin}}, \bibinfo {author} {\bibfnamefont {D.~G.}\ \bibnamefont
  {Richards}}, \ and\ \bibinfo {author} {\bibfnamefont {K.~J.}\ \bibnamefont
  {Juge}} (\bibinfo {collaboration} {Hadron Spectrum}),\ }\href {\doibase
  10.1103/PhysRevD.80.054506} {\bibfield  {journal} {\bibinfo  {journal} {Phys.
  Rev.}\ }\textbf {\bibinfo {volume} {D80}},\ \bibinfo {pages} {054506}
  (\bibinfo {year} {2009})},\ \Eprint {http://arxiv.org/abs/0905.2160}
  {arXiv:0905.2160 [hep-lat]} \BibitemShut {NoStop}%
\bibitem [{\citenamefont {Woss}\ \emph {et~al.}(2021)\citenamefont {Woss},
  \citenamefont {Dudek}, \citenamefont {Edwards}, \citenamefont {Thomas},\ and\
  \citenamefont {Wilson}}]{Woss:2020ayi}%
  \BibitemOpen
  \bibfield  {author} {\bibinfo {author} {\bibfnamefont {A.~J.}\ \bibnamefont
  {Woss}}, \bibinfo {author} {\bibfnamefont {J.~J.}\ \bibnamefont {Dudek}},
  \bibinfo {author} {\bibfnamefont {R.~G.}\ \bibnamefont {Edwards}}, \bibinfo
  {author} {\bibfnamefont {C.~E.}\ \bibnamefont {Thomas}}, \ and\ \bibinfo
  {author} {\bibfnamefont {D.~J.}\ \bibnamefont {Wilson}} (\bibinfo
  {collaboration} {Hadron Spectrum}),\ }\href {\doibase
  10.1103/PhysRevD.103.054502} {\bibfield  {journal} {\bibinfo  {journal}
  {Phys. Rev. D}\ }\textbf {\bibinfo {volume} {103}},\ \bibinfo {pages}
  {054502} (\bibinfo {year} {2021})},\ \Eprint
  {http://arxiv.org/abs/2009.10034} {arXiv:2009.10034 [hep-lat]} \BibitemShut
  {NoStop}%
\bibitem [{\citenamefont {Johnson}(1982)}]{Johnson:1982yq}%
  \BibitemOpen
  \bibfield  {author} {\bibinfo {author} {\bibfnamefont {R.~C.}\ \bibnamefont
  {Johnson}},\ }\href {\doibase 10.1016/0370-2693(82)90134-4} {\bibfield
  {journal} {\bibinfo  {journal} {Phys. Lett.}\ }\textbf {\bibinfo {volume}
  {B114}},\ \bibinfo {pages} {147} (\bibinfo {year} {1982})}\BibitemShut
  {NoStop}%
\bibitem [{\citenamefont {Moore}\ and\ \citenamefont
  {Fleming}(2006)}]{Moore:2005dw}%
  \BibitemOpen
  \bibfield  {author} {\bibinfo {author} {\bibfnamefont {D.~C.}\ \bibnamefont
  {Moore}}\ and\ \bibinfo {author} {\bibfnamefont {G.~T.}\ \bibnamefont
  {Fleming}},\ }\href {\doibase 10.1103/PhysRevD.73.014504,
  10.1103/PhysRevD.74.079905} {\bibfield  {journal} {\bibinfo  {journal} {Phys.
  Rev.}\ }\textbf {\bibinfo {volume} {D73}},\ \bibinfo {pages} {014504}
  (\bibinfo {year} {2006})},\ \bibinfo {note} {[Erratum: Phys.
  Rev.D74,079905(2006)]},\ \Eprint {http://arxiv.org/abs/hep-lat/0507018}
  {arXiv:hep-lat/0507018 [hep-lat]} \BibitemShut {NoStop}%
\bibitem [{\citenamefont {Thomas}\ \emph {et~al.}(2012)\citenamefont {Thomas},
  \citenamefont {Edwards},\ and\ \citenamefont {Dudek}}]{Thomas:2011rh}%
  \BibitemOpen
  \bibfield  {author} {\bibinfo {author} {\bibfnamefont {C.~E.}\ \bibnamefont
  {Thomas}}, \bibinfo {author} {\bibfnamefont {R.~G.}\ \bibnamefont {Edwards}},
  \ and\ \bibinfo {author} {\bibfnamefont {J.~J.}\ \bibnamefont {Dudek}},\
  }\href {\doibase 10.1103/PhysRevD.85.014507, 10.1103/PhysRevD.85.039901}
  {\bibfield  {journal} {\bibinfo  {journal} {Phys. Rev.}\ }\textbf {\bibinfo
  {volume} {D85}},\ \bibinfo {pages} {014507} (\bibinfo {year} {2012})},\
  \Eprint {http://arxiv.org/abs/1107.1930} {arXiv:1107.1930 [hep-lat]}
  \BibitemShut {NoStop}%
\bibitem [{\citenamefont {Dudek}\ \emph {et~al.}(2010)\citenamefont {Dudek},
  \citenamefont {Edwards}, \citenamefont {Peardon}, \citenamefont {Richards},\
  and\ \citenamefont {Thomas}}]{Dudek:2010wm}%
  \BibitemOpen
  \bibfield  {author} {\bibinfo {author} {\bibfnamefont {J.~J.}\ \bibnamefont
  {Dudek}}, \bibinfo {author} {\bibfnamefont {R.~G.}\ \bibnamefont {Edwards}},
  \bibinfo {author} {\bibfnamefont {M.~J.}\ \bibnamefont {Peardon}}, \bibinfo
  {author} {\bibfnamefont {D.~G.}\ \bibnamefont {Richards}}, \ and\ \bibinfo
  {author} {\bibfnamefont {C.~E.}\ \bibnamefont {Thomas}},\ }\href {\doibase
  10.1103/PhysRevD.82.034508} {\bibfield  {journal} {\bibinfo  {journal} {Phys.
  Rev.}\ }\textbf {\bibinfo {volume} {D82}},\ \bibinfo {pages} {034508}
  (\bibinfo {year} {2010})},\ \Eprint {http://arxiv.org/abs/1004.4930}
  {arXiv:1004.4930 [hep-ph]} \BibitemShut {NoStop}%
\bibitem [{\citenamefont {Jay}\ and\ \citenamefont {Neil}(2021)}]{Jay:2020jkz}%
  \BibitemOpen
  \bibfield  {author} {\bibinfo {author} {\bibfnamefont {W.~I.}\ \bibnamefont
  {Jay}}\ and\ \bibinfo {author} {\bibfnamefont {E.~T.}\ \bibnamefont {Neil}},\
  }\href {\doibase 10.1103/PhysRevD.103.114502} {\bibfield  {journal} {\bibinfo
   {journal} {Phys. Rev. D}\ }\textbf {\bibinfo {volume} {103}},\ \bibinfo
  {pages} {114502} (\bibinfo {year} {2021})},\ \Eprint
  {http://arxiv.org/abs/2008.01069} {arXiv:2008.01069 [stat.ME]} \BibitemShut
  {NoStop}%
\bibitem [{\citenamefont {Radhakrishnan}\ \emph {et~al.}(2022)\citenamefont
  {Radhakrishnan}, \citenamefont {Dudek},\ and\ \citenamefont
  {Edwards}}]{Radhakrishnan:2022ubg}%
  \BibitemOpen
  \bibfield  {author} {\bibinfo {author} {\bibfnamefont {A.}~\bibnamefont
  {Radhakrishnan}}, \bibinfo {author} {\bibfnamefont {J.~J.}\ \bibnamefont
  {Dudek}}, \ and\ \bibinfo {author} {\bibfnamefont {R.~G.}\ \bibnamefont
  {Edwards}} (\bibinfo {collaboration} {Hadron Spectrum}),\ }\href {\doibase
  10.1103/PhysRevD.106.114513} {\bibfield  {journal} {\bibinfo  {journal}
  {Phys. Rev. D}\ }\textbf {\bibinfo {volume} {106}},\ \bibinfo {pages}
  {114513} (\bibinfo {year} {2022})},\ \Eprint
  {http://arxiv.org/abs/2208.13755} {arXiv:2208.13755 [hep-lat]} \BibitemShut
  {NoStop}%
\bibitem [{\citenamefont {Guo}\ \emph {et~al.}(2013)\citenamefont {Guo},
  \citenamefont {Dudek}, \citenamefont {Edwards},\ and\ \citenamefont
  {Szczepaniak}}]{Guo:2012hv}%
  \BibitemOpen
  \bibfield  {author} {\bibinfo {author} {\bibfnamefont {P.}~\bibnamefont
  {Guo}}, \bibinfo {author} {\bibfnamefont {J.}~\bibnamefont {Dudek}}, \bibinfo
  {author} {\bibfnamefont {R.}~\bibnamefont {Edwards}}, \ and\ \bibinfo
  {author} {\bibfnamefont {A.~P.}\ \bibnamefont {Szczepaniak}},\ }\href
  {\doibase 10.1103/PhysRevD.88.014501} {\bibfield  {journal} {\bibinfo
  {journal} {Phys. Rev.}\ }\textbf {\bibinfo {volume} {D88}},\ \bibinfo {pages}
  {014501} (\bibinfo {year} {2013})},\ \Eprint {http://arxiv.org/abs/1211.0929}
  {arXiv:1211.0929 [hep-lat]} \BibitemShut {NoStop}%
\bibitem [{\citenamefont {Woss}\ \emph {et~al.}(2020)\citenamefont {Woss},
  \citenamefont {Wilson},\ and\ \citenamefont {Dudek}}]{Woss:2020cmp}%
  \BibitemOpen
  \bibfield  {author} {\bibinfo {author} {\bibfnamefont {A.~J.}\ \bibnamefont
  {Woss}}, \bibinfo {author} {\bibfnamefont {D.~J.}\ \bibnamefont {Wilson}}, \
  and\ \bibinfo {author} {\bibfnamefont {J.~J.}\ \bibnamefont {Dudek}}
  (\bibinfo {collaboration} {Hadron Spectrum}),\ }\href {\doibase
  10.1103/PhysRevD.101.114505} {\bibfield  {journal} {\bibinfo  {journal}
  {Phys. Rev. D}\ }\textbf {\bibinfo {volume} {101}},\ \bibinfo {pages}
  {114505} (\bibinfo {year} {2020})},\ \Eprint
  {http://arxiv.org/abs/2001.08474} {arXiv:2001.08474 [hep-lat]} \BibitemShut
  {NoStop}%
\bibitem [{\citenamefont {Wilson}\ \emph
  {et~al.}(2015{\natexlab{a}})\citenamefont {Wilson}, \citenamefont {Dudek},
  \citenamefont {Edwards},\ and\ \citenamefont {Thomas}}]{Wilson:2014cna}%
  \BibitemOpen
  \bibfield  {author} {\bibinfo {author} {\bibfnamefont {D.~J.}\ \bibnamefont
  {Wilson}}, \bibinfo {author} {\bibfnamefont {J.~J.}\ \bibnamefont {Dudek}},
  \bibinfo {author} {\bibfnamefont {R.~G.}\ \bibnamefont {Edwards}}, \ and\
  \bibinfo {author} {\bibfnamefont {C.~E.}\ \bibnamefont {Thomas}},\ }\href
  {\doibase 10.1103/PhysRevD.91.054008} {\bibfield  {journal} {\bibinfo
  {journal} {Phys. Rev.}\ }\textbf {\bibinfo {volume} {D91}},\ \bibinfo {pages}
  {054008} (\bibinfo {year} {2015}{\natexlab{a}})},\ \Eprint
  {http://arxiv.org/abs/1411.2004} {arXiv:1411.2004 [hep-ph]} \BibitemShut
  {NoStop}%
\bibitem [{\citenamefont {Dudek}\ \emph {et~al.}(2014)\citenamefont {Dudek},
  \citenamefont {Edwards}, \citenamefont {Thomas},\ and\ \citenamefont
  {Wilson}}]{Dudek:2014qha}%
  \BibitemOpen
  \bibfield  {author} {\bibinfo {author} {\bibfnamefont {J.~J.}\ \bibnamefont
  {Dudek}}, \bibinfo {author} {\bibfnamefont {R.~G.}\ \bibnamefont {Edwards}},
  \bibinfo {author} {\bibfnamefont {C.~E.}\ \bibnamefont {Thomas}}, \ and\
  \bibinfo {author} {\bibfnamefont {D.~J.}\ \bibnamefont {Wilson}} (\bibinfo
  {collaboration} {Hadron Spectrum}),\ }\href {\doibase
  10.1103/PhysRevLett.113.182001} {\bibfield  {journal} {\bibinfo  {journal}
  {Phys. Rev. Lett.}\ }\textbf {\bibinfo {volume} {113}},\ \bibinfo {pages}
  {182001} (\bibinfo {year} {2014})},\ \Eprint {http://arxiv.org/abs/1406.4158}
  {arXiv:1406.4158 [hep-ph]} \BibitemShut {NoStop}%
\bibitem [{\citenamefont {Wilson}\ \emph
  {et~al.}(2015{\natexlab{b}})\citenamefont {Wilson}, \citenamefont {Briceño},
  \citenamefont {Dudek}, \citenamefont {Edwards},\ and\ \citenamefont
  {Thomas}}]{Wilson:2015dqa}%
  \BibitemOpen
  \bibfield  {author} {\bibinfo {author} {\bibfnamefont {D.~J.}\ \bibnamefont
  {Wilson}}, \bibinfo {author} {\bibfnamefont {R.~A.}\ \bibnamefont
  {Briceño}}, \bibinfo {author} {\bibfnamefont {J.~J.}\ \bibnamefont {Dudek}},
  \bibinfo {author} {\bibfnamefont {R.~G.}\ \bibnamefont {Edwards}}, \ and\
  \bibinfo {author} {\bibfnamefont {C.~E.}\ \bibnamefont {Thomas}},\ }\href
  {\doibase 10.1103/PhysRevD.92.094502} {\bibfield  {journal} {\bibinfo
  {journal} {Phys. Rev.}\ }\textbf {\bibinfo {volume} {D92}},\ \bibinfo {pages}
  {094502} (\bibinfo {year} {2015}{\natexlab{b}})},\ \Eprint
  {http://arxiv.org/abs/1507.02599} {arXiv:1507.02599 [hep-ph]} \BibitemShut
  {NoStop}%
\bibitem [{\citenamefont {Moir}\ \emph {et~al.}(2016)\citenamefont {Moir},
  \citenamefont {Peardon}, \citenamefont {Ryan}, \citenamefont {Thomas},\ and\
  \citenamefont {Wilson}}]{Moir:2016srx}%
  \BibitemOpen
  \bibfield  {author} {\bibinfo {author} {\bibfnamefont {G.}~\bibnamefont
  {Moir}}, \bibinfo {author} {\bibfnamefont {M.}~\bibnamefont {Peardon}},
  \bibinfo {author} {\bibfnamefont {S.~M.}\ \bibnamefont {Ryan}}, \bibinfo
  {author} {\bibfnamefont {C.~E.}\ \bibnamefont {Thomas}}, \ and\ \bibinfo
  {author} {\bibfnamefont {D.~J.}\ \bibnamefont {Wilson}},\ }\href {\doibase
  10.1007/JHEP10(2016)011} {\bibfield  {journal} {\bibinfo  {journal} {JHEP}\
  }\textbf {\bibinfo {volume} {10}},\ \bibinfo {pages} {011} (\bibinfo {year}
  {2016})},\ \Eprint {http://arxiv.org/abs/1607.07093} {arXiv:1607.07093
  [hep-lat]} \BibitemShut {NoStop}%
\bibitem [{\citenamefont {Dudek}\ \emph {et~al.}(2016)\citenamefont {Dudek},
  \citenamefont {Edwards},\ and\ \citenamefont {Wilson}}]{Dudek:2016cru}%
  \BibitemOpen
  \bibfield  {author} {\bibinfo {author} {\bibfnamefont {J.~J.}\ \bibnamefont
  {Dudek}}, \bibinfo {author} {\bibfnamefont {R.~G.}\ \bibnamefont {Edwards}},
  \ and\ \bibinfo {author} {\bibfnamefont {D.~J.}\ \bibnamefont {Wilson}}
  (\bibinfo {collaboration} {Hadron Spectrum}),\ }\href {\doibase
  10.1103/PhysRevD.93.094506} {\bibfield  {journal} {\bibinfo  {journal} {Phys.
  Rev.}\ }\textbf {\bibinfo {volume} {D93}},\ \bibinfo {pages} {094506}
  (\bibinfo {year} {2016})},\ \Eprint {http://arxiv.org/abs/1602.05122}
  {arXiv:1602.05122 [hep-ph]} \BibitemShut {NoStop}%
\bibitem [{\citenamefont {Briceno}\ \emph
  {et~al.}(2018{\natexlab{b}})\citenamefont {Briceno}, \citenamefont {Dudek},
  \citenamefont {Edwards},\ and\ \citenamefont {Wilson}}]{Briceno:2017qmb}%
  \BibitemOpen
  \bibfield  {author} {\bibinfo {author} {\bibfnamefont {R.~A.}\ \bibnamefont
  {Briceno}}, \bibinfo {author} {\bibfnamefont {J.~J.}\ \bibnamefont {Dudek}},
  \bibinfo {author} {\bibfnamefont {R.~G.}\ \bibnamefont {Edwards}}, \ and\
  \bibinfo {author} {\bibfnamefont {D.~J.}\ \bibnamefont {Wilson}},\ }\href
  {\doibase 10.1103/PhysRevD.97.054513} {\bibfield  {journal} {\bibinfo
  {journal} {Phys. Rev.}\ }\textbf {\bibinfo {volume} {D97}},\ \bibinfo {pages}
  {054513} (\bibinfo {year} {2018}{\natexlab{b}})},\ \Eprint
  {http://arxiv.org/abs/1708.06667} {arXiv:1708.06667 [hep-lat]} \BibitemShut
  {NoStop}%
\bibitem [{\citenamefont {Woss}\ \emph {et~al.}(2019)\citenamefont {Woss},
  \citenamefont {Thomas}, \citenamefont {Dudek}, \citenamefont {Edwards},\ and\
  \citenamefont {Wilson}}]{Woss:2019hse}%
  \BibitemOpen
  \bibfield  {author} {\bibinfo {author} {\bibfnamefont {A.~J.}\ \bibnamefont
  {Woss}}, \bibinfo {author} {\bibfnamefont {C.~E.}\ \bibnamefont {Thomas}},
  \bibinfo {author} {\bibfnamefont {J.~J.}\ \bibnamefont {Dudek}}, \bibinfo
  {author} {\bibfnamefont {R.~G.}\ \bibnamefont {Edwards}}, \ and\ \bibinfo
  {author} {\bibfnamefont {D.~J.}\ \bibnamefont {Wilson}},\ }\href {\doibase
  10.1103/PhysRevD.100.054506} {\bibfield  {journal} {\bibinfo  {journal}
  {Phys. Rev.}\ }\textbf {\bibinfo {volume} {D100}},\ \bibinfo {pages} {054506}
  (\bibinfo {year} {2019})},\ \Eprint {http://arxiv.org/abs/1904.04136}
  {arXiv:1904.04136 [hep-lat]} \BibitemShut {NoStop}%
\bibitem [{\citenamefont {Prelovsek}\ \emph {et~al.}(2021)\citenamefont
  {Prelovsek}, \citenamefont {Collins}, \citenamefont {Mohler}, \citenamefont
  {Padmanath},\ and\ \citenamefont {Piemonte}}]{Prelovsek:2020eiw}%
  \BibitemOpen
  \bibfield  {author} {\bibinfo {author} {\bibfnamefont {S.}~\bibnamefont
  {Prelovsek}}, \bibinfo {author} {\bibfnamefont {S.}~\bibnamefont {Collins}},
  \bibinfo {author} {\bibfnamefont {D.}~\bibnamefont {Mohler}}, \bibinfo
  {author} {\bibfnamefont {M.}~\bibnamefont {Padmanath}}, \ and\ \bibinfo
  {author} {\bibfnamefont {S.}~\bibnamefont {Piemonte}},\ }\href {\doibase
  10.1007/JHEP06(2021)035} {\bibfield  {journal} {\bibinfo  {journal} {JHEP}\
  }\textbf {\bibinfo {volume} {06}},\ \bibinfo {pages} {035} (\bibinfo {year}
  {2021})},\ \Eprint {http://arxiv.org/abs/2011.02542} {arXiv:2011.02542
  [hep-lat]} \BibitemShut {NoStop}%
\bibitem [{\citenamefont {Lang}\ and\ \citenamefont
  {Wilson}(2022)}]{Lang:2022elg}%
  \BibitemOpen
  \bibfield  {author} {\bibinfo {author} {\bibfnamefont {N.}~\bibnamefont
  {Lang}}\ and\ \bibinfo {author} {\bibfnamefont {D.~J.}\ \bibnamefont
  {Wilson}} (\bibinfo {collaboration} {Hadron Spectrum}),\ }\href {\doibase
  10.1103/PhysRevLett.129.252001} {\bibfield  {journal} {\bibinfo  {journal}
  {Phys. Rev. Lett.}\ }\textbf {\bibinfo {volume} {129}},\ \bibinfo {pages}
  {252001} (\bibinfo {year} {2022})},\ \Eprint
  {http://arxiv.org/abs/2205.05026} {arXiv:2205.05026 [hep-ph]} \BibitemShut
  {NoStop}%
\bibitem [{\citenamefont {Bulava}\ \emph {et~al.}(2024)\citenamefont {Bulava}
  \emph {et~al.}}]{BaryonScatteringBaSc:2023ori}%
  \BibitemOpen
  \bibfield  {author} {\bibinfo {author} {\bibfnamefont {J.}~\bibnamefont
  {Bulava}} \emph {et~al.} (\bibinfo {collaboration} {Baryon Scattering
  (BaSc)}),\ }\href {\doibase 10.1103/PhysRevD.109.014511} {\bibfield
  {journal} {\bibinfo  {journal} {Phys. Rev. D}\ }\textbf {\bibinfo {volume}
  {109}},\ \bibinfo {pages} {014511} (\bibinfo {year} {2024})},\ \Eprint
  {http://arxiv.org/abs/2307.13471} {arXiv:2307.13471 [hep-lat]} \BibitemShut
  {NoStop}%
\bibitem [{\citenamefont {Wilson}\ \emph
  {et~al.}(2024{\natexlab{a}})\citenamefont {Wilson}, \citenamefont {Thomas},
  \citenamefont {Dudek},\ and\ \citenamefont
  {Edwards}}]{PhysRevLett.132.241901}%
  \BibitemOpen
  \bibfield  {author} {\bibinfo {author} {\bibfnamefont {D.~J.}\ \bibnamefont
  {Wilson}}, \bibinfo {author} {\bibfnamefont {C.~E.}\ \bibnamefont {Thomas}},
  \bibinfo {author} {\bibfnamefont {J.~J.}\ \bibnamefont {Dudek}}, \ and\
  \bibinfo {author} {\bibfnamefont {R.~G.}\ \bibnamefont {Edwards}} (\bibinfo
  {collaboration} {for the Hadron Spectrum Collaboration}),\ }\href {\doibase
  10.1103/PhysRevLett.132.241901} {\bibfield  {journal} {\bibinfo  {journal}
  {Phys. Rev. Lett.}\ }\textbf {\bibinfo {volume} {132}},\ \bibinfo {pages}
  {241901} (\bibinfo {year} {2024}{\natexlab{a}})}\BibitemShut {NoStop}%
\bibitem [{\citenamefont {Wilson}\ \emph
  {et~al.}(2024{\natexlab{b}})\citenamefont {Wilson}, \citenamefont {Thomas},
  \citenamefont {Dudek},\ and\ \citenamefont {Edwards}}]{PhysRevD.109.114503}%
  \BibitemOpen
  \bibfield  {author} {\bibinfo {author} {\bibfnamefont {D.~J.}\ \bibnamefont
  {Wilson}}, \bibinfo {author} {\bibfnamefont {C.~E.}\ \bibnamefont {Thomas}},
  \bibinfo {author} {\bibfnamefont {J.~J.}\ \bibnamefont {Dudek}}, \ and\
  \bibinfo {author} {\bibfnamefont {R.~G.}\ \bibnamefont {Edwards}} (\bibinfo
  {collaboration} {for the Hadron Spectrum Collaboration}),\ }\href {\doibase
  10.1103/PhysRevD.109.114503} {\bibfield  {journal} {\bibinfo  {journal}
  {Phys. Rev. D}\ }\textbf {\bibinfo {volume} {109}},\ \bibinfo {pages}
  {114503} (\bibinfo {year} {2024}{\natexlab{b}})}\BibitemShut {NoStop}%
\bibitem [{\citenamefont {Workman}\ and\ \citenamefont
  {Others}(2022)}]{Workman:2022ynf}%
  \BibitemOpen
  \bibfield  {author} {\bibinfo {author} {\bibfnamefont {R.~L.}\ \bibnamefont
  {Workman}}\ and\ \bibinfo {author} {\bibnamefont {Others}} (\bibinfo
  {collaboration} {Particle Data Group}),\ }\href {\doibase
  10.1093/ptep/ptac097} {\bibfield  {journal} {\bibinfo  {journal} {PTEP}\
  }\textbf {\bibinfo {volume} {2022}},\ \bibinfo {pages} {083C01} (\bibinfo
  {year} {2022})}\BibitemShut {NoStop}%
\bibitem [{\citenamefont {Dudek}\ \emph {et~al.}(2009)\citenamefont {Dudek},
  \citenamefont {Edwards}, \citenamefont {Peardon}, \citenamefont {Richards},\
  and\ \citenamefont {Thomas}}]{Dudek:2009qf}%
  \BibitemOpen
  \bibfield  {author} {\bibinfo {author} {\bibfnamefont {J.~J.}\ \bibnamefont
  {Dudek}}, \bibinfo {author} {\bibfnamefont {R.~G.}\ \bibnamefont {Edwards}},
  \bibinfo {author} {\bibfnamefont {M.~J.}\ \bibnamefont {Peardon}}, \bibinfo
  {author} {\bibfnamefont {D.~G.}\ \bibnamefont {Richards}}, \ and\ \bibinfo
  {author} {\bibfnamefont {C.~E.}\ \bibnamefont {Thomas}},\ }\href {\doibase
  10.1103/PhysRevLett.103.262001} {\bibfield  {journal} {\bibinfo  {journal}
  {Phys. Rev. Lett.}\ }\textbf {\bibinfo {volume} {103}},\ \bibinfo {pages}
  {262001} (\bibinfo {year} {2009})},\ \Eprint {http://arxiv.org/abs/0909.0200}
  {arXiv:0909.0200 [hep-ph]} \BibitemShut {NoStop}%
\bibitem [{\citenamefont {Dudek}(2011)}]{Dudek:2011bn}%
  \BibitemOpen
  \bibfield  {author} {\bibinfo {author} {\bibfnamefont {J.~J.}\ \bibnamefont
  {Dudek}},\ }\href {\doibase 10.1103/PhysRevD.84.074023} {\bibfield  {journal}
  {\bibinfo  {journal} {Phys. Rev.}\ }\textbf {\bibinfo {volume} {D84}},\
  \bibinfo {pages} {074023} (\bibinfo {year} {2011})},\ \Eprint
  {http://arxiv.org/abs/1106.5515} {arXiv:1106.5515 [hep-ph]} \BibitemShut
  {NoStop}%
\bibitem [{\citenamefont {Dudek}\ \emph {et~al.}(2013)\citenamefont {Dudek},
  \citenamefont {Edwards}, \citenamefont {Guo},\ and\ \citenamefont
  {Thomas}}]{Dudek:2013yja}%
  \BibitemOpen
  \bibfield  {author} {\bibinfo {author} {\bibfnamefont {J.~J.}\ \bibnamefont
  {Dudek}}, \bibinfo {author} {\bibfnamefont {R.~G.}\ \bibnamefont {Edwards}},
  \bibinfo {author} {\bibfnamefont {P.}~\bibnamefont {Guo}}, \ and\ \bibinfo
  {author} {\bibfnamefont {C.~E.}\ \bibnamefont {Thomas}} (\bibinfo
  {collaboration} {Hadron Spectrum}),\ }\href {\doibase
  10.1103/PhysRevD.88.094505} {\bibfield  {journal} {\bibinfo  {journal} {Phys.
  Rev.}\ }\textbf {\bibinfo {volume} {D88}},\ \bibinfo {pages} {094505}
  (\bibinfo {year} {2013})},\ \Eprint {http://arxiv.org/abs/1309.2608}
  {arXiv:1309.2608 [hep-lat]} \BibitemShut {NoStop}%
\bibitem [{\citenamefont {de~Swart}(1963)}]{deSwart:1963pdg}%
  \BibitemOpen
  \bibfield  {author} {\bibinfo {author} {\bibfnamefont {J.~J.}\ \bibnamefont
  {de~Swart}},\ }\href {\doibase 10.1103/RevModPhys.35.916} {\bibfield
  {journal} {\bibinfo  {journal} {Rev. Mod. Phys.}\ }\textbf {\bibinfo {volume}
  {35}},\ \bibinfo {pages} {916} (\bibinfo {year} {1963})},\ \bibinfo {note}
  {[Erratum: Rev. Mod. Phys.37,326(1965)]}\BibitemShut {NoStop}%
\bibitem [{\citenamefont {Clegg}\ and\ \citenamefont
  {Donnachie}(1994)}]{Clegg:1993mt}%
  \BibitemOpen
  \bibfield  {author} {\bibinfo {author} {\bibfnamefont {A.~B.}\ \bibnamefont
  {Clegg}}\ and\ \bibinfo {author} {\bibfnamefont {A.}~\bibnamefont
  {Donnachie}},\ }\href {\doibase 10.1007/BF01555905} {\bibfield  {journal}
  {\bibinfo  {journal} {Z. Phys. C}\ }\textbf {\bibinfo {volume} {62}},\
  \bibinfo {pages} {455} (\bibinfo {year} {1994})}\BibitemShut {NoStop}%
\bibitem [{\citenamefont {Aston}\ \emph {et~al.}(1993)\citenamefont {Aston}
  \emph {et~al.}}]{Aston:1993qc}%
  \BibitemOpen
  \bibfield  {author} {\bibinfo {author} {\bibfnamefont {D.}~\bibnamefont
  {Aston}} \emph {et~al.},\ }\href {\doibase 10.1016/0370-2693(93)90620-W}
  {\bibfield  {journal} {\bibinfo  {journal} {Phys. Lett. B}\ }\textbf
  {\bibinfo {volume} {308}},\ \bibinfo {pages} {186} (\bibinfo {year}
  {1993})}\BibitemShut {NoStop}%
\bibitem [{\citenamefont {Aubert}\ \emph {et~al.}(2008)\citenamefont {Aubert}
  \emph {et~al.}}]{BaBar:2007ceh}%
  \BibitemOpen
  \bibfield  {author} {\bibinfo {author} {\bibfnamefont {B.}~\bibnamefont
  {Aubert}} \emph {et~al.} (\bibinfo {collaboration} {BaBar}),\ }\href
  {\doibase 10.1103/PhysRevD.77.092002} {\bibfield  {journal} {\bibinfo
  {journal} {Phys. Rev. D}\ }\textbf {\bibinfo {volume} {77}},\ \bibinfo
  {pages} {092002} (\bibinfo {year} {2008})},\ \Eprint
  {http://arxiv.org/abs/0710.4451} {arXiv:0710.4451 [hep-ex]} \BibitemShut
  {NoStop}%
\bibitem [{\citenamefont {Hansen}\ \emph {et~al.}(2021)\citenamefont {Hansen},
  \citenamefont {Brice\~no}, \citenamefont {Edwards}, \citenamefont {Thomas},\
  and\ \citenamefont {Wilson}}]{Hansen:2020otl}%
  \BibitemOpen
  \bibfield  {author} {\bibinfo {author} {\bibfnamefont {M.~T.}\ \bibnamefont
  {Hansen}}, \bibinfo {author} {\bibfnamefont {R.~A.}\ \bibnamefont
  {Brice\~no}}, \bibinfo {author} {\bibfnamefont {R.~G.}\ \bibnamefont
  {Edwards}}, \bibinfo {author} {\bibfnamefont {C.~E.}\ \bibnamefont {Thomas}},
  \ and\ \bibinfo {author} {\bibfnamefont {D.~J.}\ \bibnamefont {Wilson}}
  (\bibinfo {collaboration} {Hadron Spectrum}),\ }\href {\doibase
  10.1103/PhysRevLett.126.012001} {\bibfield  {journal} {\bibinfo  {journal}
  {Phys. Rev. Lett.}\ }\textbf {\bibinfo {volume} {126}},\ \bibinfo {pages}
  {012001} (\bibinfo {year} {2021})},\ \Eprint
  {http://arxiv.org/abs/2009.04931} {arXiv:2009.04931 [hep-lat]} \BibitemShut
  {NoStop}%
\bibitem [{\citenamefont {Edwards}\ and\ \citenamefont
  {Joo}(2005)}]{Edwards:2004sx}%
  \BibitemOpen
  \bibfield  {author} {\bibinfo {author} {\bibfnamefont {R.~G.}\ \bibnamefont
  {Edwards}}\ and\ \bibinfo {author} {\bibfnamefont {B.}~\bibnamefont {Joo}}
  (\bibinfo {collaboration} {SciDAC, LHPC, UKQCD}),\ }\bibfield  {booktitle}
  {\emph {\bibinfo {booktitle} {{Lattice field theory. Proceedings, 22nd
  International Symposium, Lattice 2004, Batavia, USA, June 21-26, 2004}}},\
  }\href {\doibase 10.1016/j.nuclphysbps.2004.11.254} {\bibfield  {journal}
  {\bibinfo  {journal} {Nucl. Phys. Proc. Suppl.}\ }\textbf {\bibinfo {volume}
  {140}},\ \bibinfo {pages} {832} (\bibinfo {year} {2005})},\ \Eprint
  {http://arxiv.org/abs/hep-lat/0409003} {arXiv:hep-lat/0409003 [hep-lat]}
  \BibitemShut {NoStop}%
\bibitem [{\citenamefont {Clark}\ \emph {et~al.}(2010)\citenamefont {Clark},
  \citenamefont {Babich}, \citenamefont {Barros}, \citenamefont {Brower},\ and\
  \citenamefont {Rebbi}}]{Clark:2009wm}%
  \BibitemOpen
  \bibfield  {author} {\bibinfo {author} {\bibfnamefont {M.~A.}\ \bibnamefont
  {Clark}}, \bibinfo {author} {\bibfnamefont {R.}~\bibnamefont {Babich}},
  \bibinfo {author} {\bibfnamefont {K.}~\bibnamefont {Barros}}, \bibinfo
  {author} {\bibfnamefont {R.~C.}\ \bibnamefont {Brower}}, \ and\ \bibinfo
  {author} {\bibfnamefont {C.}~\bibnamefont {Rebbi}},\ }\href {\doibase
  10.1016/j.cpc.2010.05.002} {\bibfield  {journal} {\bibinfo  {journal}
  {Comput. Phys. Commun.}\ }\textbf {\bibinfo {volume} {181}},\ \bibinfo
  {pages} {1517} (\bibinfo {year} {2010})},\ \Eprint
  {http://arxiv.org/abs/0911.3191} {arXiv:0911.3191 [hep-lat]} \BibitemShut
  {NoStop}%
\bibitem [{\citenamefont {Babich}\ \emph {et~al.}(2010)\citenamefont {Babich},
  \citenamefont {Clark},\ and\ \citenamefont {Joo}}]{Babich:2010mu}%
  \BibitemOpen
  \bibfield  {author} {\bibinfo {author} {\bibfnamefont {R.}~\bibnamefont
  {Babich}}, \bibinfo {author} {\bibfnamefont {M.~A.}\ \bibnamefont {Clark}}, \
  and\ \bibinfo {author} {\bibfnamefont {B.}~\bibnamefont {Joo}},\ }in\ \href
  {http://www1.jlab.org/Ul/publications/view_pub.cfm?pub_id=10186} {\emph
  {\bibinfo {booktitle} {{SC 10 (Supercomputing 2010) New Orleans, Louisiana,
  November 13-19, 2010}}}}\ (\bibinfo {year} {2010})\ \Eprint
  {http://arxiv.org/abs/1011.0024} {arXiv:1011.0024 [hep-lat]} \BibitemShut
  {NoStop}%
\bibitem [{\citenamefont {Clark}\ \emph {et~al.}(2016)\citenamefont {Clark},
  \citenamefont {Joó}, \citenamefont {Strelchenko}, \citenamefont {Cheng},
  \citenamefont {Gambhir},\ and\ \citenamefont {Brower}}]{Clark:2016rdz}%
  \BibitemOpen
  \bibfield  {author} {\bibinfo {author} {\bibfnamefont {M.}~\bibnamefont
  {Clark}}, \bibinfo {author} {\bibfnamefont {B.}~\bibnamefont {Joó}},
  \bibinfo {author} {\bibfnamefont {A.}~\bibnamefont {Strelchenko}}, \bibinfo
  {author} {\bibfnamefont {M.}~\bibnamefont {Cheng}}, \bibinfo {author}
  {\bibfnamefont {A.}~\bibnamefont {Gambhir}}, \ and\ \bibinfo {author}
  {\bibfnamefont {R.}~\bibnamefont {Brower}},\ }in\ \href@noop {} {\emph
  {\bibinfo {booktitle} {SC '16: Proceedings of the International Conference
  for High Performance Computing, Networking, Storage and Analysis}}}\
  (\bibinfo {year} {2016})\ pp.\ \bibinfo {pages} {{795--806}},\ \Eprint
  {http://arxiv.org/abs/1612.07873} {arXiv:1612.07873 [hep-lat]} \BibitemShut
  {NoStop}%
\end{thebibliography}%


\end{document}